# Integrative Experiments Identify How Punishment Impacts Welfare in Public Goods Games

Mohammed Alsobay[a], David G. Rand[a,b], Duncan J. Watts[c,d,e], and Abdullah Almaatouq[a,b,*]

[a] Sloan School of Management, Massachusetts Institute of Technology
[b] Institute for Data, Systems, and Society, Massachusetts Institute of Technology
[c] The Wharton School, University of Pennsylvania
[d] Department of Computer and Information Science, University of Pennsylvania
[e] Annenberg School of Communication, University of Pennsylvania

*Correspondence to: amaatouq@mit.edu

**Abstract:** Punishment as a mechanism for promoting cooperation has been studied extensively for more than two decades, but its effectiveness remains a matter of dispute. Here, we examine how punishment's impact varies across cooperative settings through a large-scale integrative experiment. We vary 14 parameters that characterize public goods games, sampling 360 experimental conditions and collecting 147,618 decisions from 7,100 participants. Our results reveal striking heterogeneity in punishment effectiveness: while punishment consistently increases contributions, its impact on payoffs (i.e., efficiency) ranges from dramatically enhancing welfare (up to 43% improvement) to severely undermining it (up to 44% reduction) depending on the cooperative context. To characterize these patterns, we developed models that outperformed human forecasters (laypeople and domain experts) in predicting punishment outcomes in new experiments. Communication emerged as the most predictive feature, followed by contribution framing (opt-out vs. opt-in), contribution type (variable vs. all-or-nothing), game length (number of rounds), peer outcome visibility (whether participants can see others' earnings), and the availability of a reward mechanism. Interestingly, however, most of these features interact to influence punishment effectiveness rather than operating independently. For example, the extent to which longer games increase the effectiveness of punishment depends on whether groups can communicate. Together, our results refocus the debate over punishment from whether or not it "works" to the specific conditions under which it does and does not work. More broadly, our study demonstrates how integrative experiments can be combined with machine learning to uncover generalizable patterns, potentially involving interactions between multiple features, and help generate novel explanations in complex social phenomena.

# Introduction

Under a wide range of circumstances humans face social (or cooperative) dilemmas in which the attainment and maintenance of cooperation (defined as choosing collective over individual benefit) is a persistent challenge (Bor et al., 2023; Johnson et al., 2020; Milinski et al., 2006; Ostrom, 1990; van Baal et al., 2022). Public goods games (PGGs) provide a stylized framework that captures the essential features of real-world cooperation problems while remaining amenable to theoretical modeling and controlled experiments (Isaac & Walker, 1988; Rand & Nowak, 2013). In a typical PGG, individuals must decide how much to contribute to a shared project that benefits everyone in the group. Contributions are multiplied by a factor ($M$), and the resulting amount is shared equally among all $N$ group members; thus, each person receives M/N per unit contributed to the project—often called the marginal per capita return (*MPCR*). When the *MPCR* is less than 1 ($M < N$), individuals face the classic social dilemma: while the group as a whole benefits most when everyone contributes fully, each individual can maximize their personal gain by contributing nothing (see Figure 1.A). The failure to cooperate, where rational pursuit of self-interest undermines collective welfare, leads to what is known as the "free-rider problem" or the "tragedy of the commons" (Hardin, 1968; Marwell & Ames, 1980; Olson, 1971).

Costly peer punishment, which allows individuals to impose penalties on peers based on their actions in the PGG, has been widely studied over the past 25 years as a way to promote cooperation in social dilemmas. Early studies demonstrated that the ability to punish effectively deters free-riding and increases cooperation in terms of contributions to the public good (Bowles, 2004; Boyd et al., 2003; Fehr & Gächter, 1999, 2002; Henrich et al., 2006). However, because punishment is costly for both the punisher and the punished, higher contribution levels to the public good do not necessarily translate to improved collective welfare. Several studies found that punishment's costs can outweigh its benefits in promoting cooperation, leading to reduced overall welfare (Dreber et al., 2008; Herrmann et al., 2008; Wu et al., 2009). Subsequent research examined how various contextual factors influence whether increased cooperation outweighs punishment costs. These factors include structural parameters like the number of interaction rounds (Gächter et al., 2008), group size (Carpenter, 2007; Xu et al., 2013), and payoff parameters (Gächter et al., 2024); the effectiveness of the punishment technology (Egas & Riedl, 2008; Nikiforakis & Normann, 2008); the existence of other mechanisms like reward (Rand et al., 2009); and social dynamics, including the ability to communicate (Bochet et al., 2006); the type of peer information available (Nikiforakis, 2010), the benefits of reputation (Raihani & Bshary, 2015; Santos et al., 2013), and the ability to coordinate against free-riders (Boyd et al., 2010; Molleman et al., 2019).

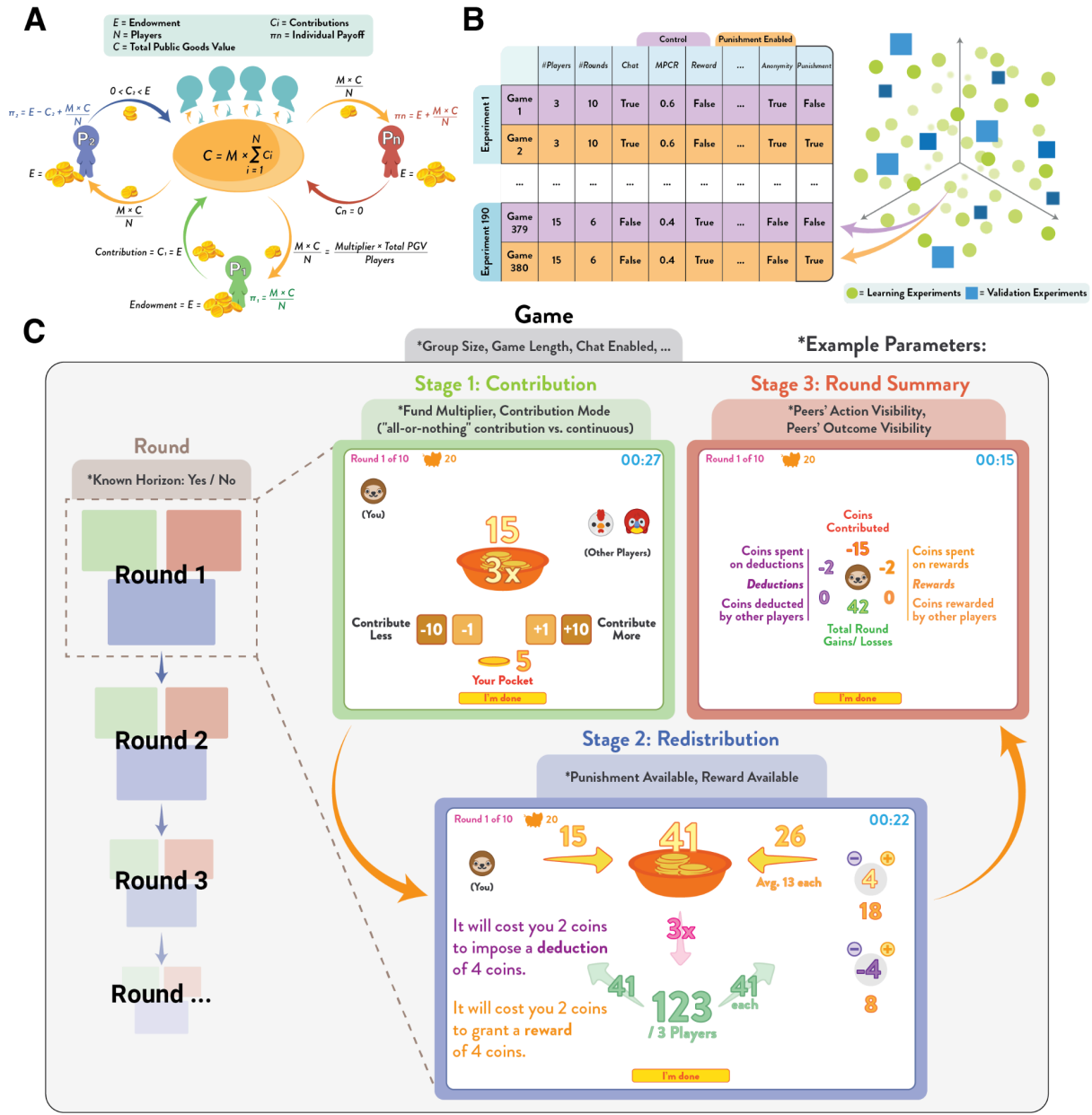

**Figure 1. Experimental design and implementation.** Panel (A) depicts a schematic representation of a typical public goods game, illustrating the basic tension between individual and collective interests. Panel (B) shows how each PGG setting is defined by a 14-dimensional parameter vector, with pairs of games—treatment (punishment enabled) and control (punishment disabled)—sharing identical parameters except for punishment availability. Data collection occurred in two waves: Wave 1 (learning experiments) used a Sobol sequence to systematically sample 320 experimental conditions with one realization each, while Wave 2 (validation experiments) randomly selected 40 new experimental conditions with multiple realizations each, providing higher precision but lower breadth. Panel (C) shows the interface from a participant's perspective, progressing through contribution, distribution, and summary stages. Interface elements (e.g., chat functionality, outcome visibility, punishment/reward options) correspond to varying PGG parameters across games. See Methods and SI for details on pre-registration, sampling strategy, and interface implementation.

Despite this large body of research, the specific conditions under which punishment actually improves group welfare remain unclear. We argue that this lack of clarity ultimately derives from how experimental studies are typically conducted—by manipulating only one or a few theoretically informed factors in isolation. For instance, a study might vary the time horizon while holding constant group size, payoff structure, cost of punishment, and time pressure, assuming these to be inconsequential "nuisance" parameters. Although standard, this "one at a time" approach comes with at least two important limitations (Almaatouq et al., 2022; Almaatouq, Griffiths, et al., 2024). First, by manipulating factors individually, such studies cannot compare the relative importance of different contextual factors. For example, we cannot determine whether the number of rounds has a greater impact on punishment effectiveness than does the strength of the punishment mechanism. Even meta-analyses cannot address this problem because studies differ in unspecified ways (experimental procedures, PGG parameter values, participant populations, etc.), making their results incommensurable, and publication bias (the "file-drawer problem") may systematically exclude certain experimental conditions—such as those showing null or negative effects. Second, this approach may blind us to higher-order, untheorized interactions between factors. The effect of punishment strength, for instance, might depend on the game length, group size, and existence of reward mechanisms in ways that single-factor studies cannot capture.

In this paper, we address both issues through a large-scale integrative experiment (Almaatouq et al., 2022) that provides a more comprehensive answer about the conditions under which punishment effectively promotes cooperation. Building on the extensive literature that has identified numerous factors affecting punishment effectiveness, we take the next step: quantifying how these factors work together to determine outcomes. The key insight is that knowing which factors matter is fundamentally different from knowing how much each matters and how they interact (Almaatouq et al., 2022; Hofman et al., 2021; Newell, 1973). This difference is crucial for answering the practical question of whether implementing punishment in a given context will help or harm cooperation.

The integrative approach, therefore, shifts the scientific question from "does this effect exist?" to "how much does it matter relative to other effects, and when?" This enables us to move beyond documenting individual effects to predicting outcomes when multiple factors operate simultaneously (addressing the practical question of when to implement punishment mechanisms). In this study, we simultaneously varied 14 parameters (see Table 1) that describe the design space of PGGs, encompassing fundamental game structure (group size, game length, contribution mechanism), communication and information structure (availability of communication, visibility of peer actions and outcomes, knowledge of the game horizon), and incentive structure (marginal per capita return, punishment and reward mechanisms—including their availability, cost, and effectiveness). Each PGG scenario is represented as a point in this 14-dimensional space, which allows us to quantify similarities and differences between scenarios in terms of these potentially relevant attributes. We then procedurally sampled experiments from this design space. Each experiment comprised a pair of games—one with punishment enabled (treatment) and one without (control)—that otherwise shared identical parameter values.

We began with a set of learning experiments, in which we used a Sobol sequence, a quasi-random strategy that fills the space of possibilities evenly (Joe & Kuo, 2008; Owen, 1998; Virtanen et al., 2020), to select 320 experimental conditions. Consistent with recent related methodological innovations and experimental design principles showing that, for a fixed overall sample size (i.e., fixed budget), estimation accuracy for models with interactions is maximized by sampling many unique design points rather than repeated observations (Baribault et al., 2018; DeKay et al., 2022; Kenny & Judd, 2019; Moerbeek & Teerenstra, 2021), each experiment was conducted once, offering a dataset with a high breadth of game scenarios but low precision. This data collection wave provides a dataset with high breadth across the parameter space, allowing us to explore and uncover empirical patterns between cooperative settings and outcomes. We then conducted validation experiments (Wave 2), which served as our confirmatory phase. Using pre-registered conditions and sample sizes (see Methods & SI), we selected 20 new, randomly chosen experiments (40 experimental conditions), each with 8–12 realizations/trials. While Wave 1 prioritized breadth for pattern discovery, Wave 2 required higher precision for evaluating specific predictions with sufficient statistical power. This two-phase approach mirrors best practices in theory development by first identifying patterns and distilling them into a model, then rigorously testing whether that model's predictions and implications hold in completely new data, directly addressing concerns about spurious patterns (i.e., overfitting) that might not generalize beyond the exploratory data (Rocca & Yarkoni, 2021; Yarkoni & Westfall, 2017). Figure 1.B illustrates the overall design.

Across both experiments, we sampled 360 unique experimental conditions, yielding 147,618 decisions (102,343 contributions, 15,471 punishments, and 29,804 rewards) from 7,100 participants. Participants were recruited from two online platforms for research participants: Amazon Mechanical Turk (via a curated panel) and Prolific. The experiments were implemented using the Empirica platform (Almaatouq et al., 2021), an open-source software for integrative experimentation through multi-participant online games with real-time interactivity. Additional details on study design, participant recruitment, and the apparatus are in the Methods section.

In terms of scale (e.g., number of participants, number of decisions), our experiment is among the largest single experiments to date, while in terms of scope (i.e., number of conditions tested), it is comparable to the largest meta-analyses in the punishment and sanctioning literature (Balliet et al., 2011; Balliet & Van Lange, 2013; Jin et al., 2024). Beyond scale and scope, our integrative approach also has several key advantages over traditional studies and meta-analyses: (1) all factors were manipulated systematically and simultaneously, allowing us to measure the relative importance of various contextual parameters' impacts and characterize higher-order interactions between these parameters; (2) all experiments were run under similar protocols (e.g., participant recruitment, user interface), enabling more direct comparisons and minimizing the risk of "hidden moderators"; (3) we avoided publication bias by treating all sampled conditions as informative regardless of statistical significance; (4) all inferences were evaluated on the pre-registered Wave 2 validation experiments (i.e., the out-of-sample dataset), ensuring the robustness of our findings.

**Table 1. Design space parameters**

| Parameter | Description | Values |
|---|---|---|
| Group Size | Number of players in the game | 2-20 players |
| Game Length | Number of rounds in the game | 1-30 rounds |
| Contribution Type | Whether players can contribute any amount or must choose between contributing all or none of their endowment | {variable, all-or-nothing} |
| Contribution Framing | Whether each player's endowment starts in their private account (opt-in) or in the public fund (opt-out) | {opt-in, opt-out} |
| MPCR | Marginal per capita return on contributions, defined as the fund multiplier divided by group size | 0.06-0.7 |
| Communication | Whether players can communicate through a chat window during the game | {enabled, disabled} |
| Peer Outcome Visibility | Whether players can see summaries of others' earnings and punishments/rewards received in each round | {visible, hidden} |
| Actor Anonymity | Whether the identity of players who punish or reward others is revealed | {revealed, hidden} |
| Horizon Knowledge | Whether players are shown the total number of rounds and rounds remaining | {known, unknown} |
| Punishment | Whether players can impose costly penalties on others. This is the focal treatment | {enabled, disabled} |
| Peer Incentive Cost | Number of coins a player must spend to impose one unit of punishment or grant one unit of reward | 1-4 coins |
| Punishment Impact | Number of coins deducted from the punished player per coin spent punishing | 1-4 coins |
| Reward | Whether players can grant costly rewards to others | {enabled, disabled} |
| Reward Impact | Number of coins granted to the rewarded player per coin spent rewarding | 0.5-1.5 coins |

# Results

**Heterogeneous effects of punishment across experimental conditions.** We begin by considering the effect of punishment on cooperation—the per round amount contributed to the public good. As shown in Figure 2A, on average, punishment increased contributions from 73% to 80% of the endowment (coefficient = 0.066, SE = 0.018, 95% CI [0.029, 0.103], p < 0.001, N = 53,973 individual-level contribution decisions), representing a 7 percentage point increase in contribution levels (or a 9.6% relative increase in contributions). This overall positive effect on contributions was replicated in our validation experiments where, on average, punishment increased contributions from 74% to 82% of endowment (coefficient = 0.077, SE = 0.016, 95% CI [0.045, 0.108], p < 0.001, N = 48,370 individual-level contribution decisions), and aligns with earlier studies that demonstrated punishment's ability to deter free-riding and to increase contributions (Balliet et al., 2011; Bowles, 2004; Boyd et al., 2003; Fehr & Gächter, 1999, 2002; Henrich et al., 2006). To account for the hierarchical structure of the data, the effect of punishment on contributions is estimated in a mixed-effects linear model with random intercepts at the participant and group levels; for further details on regression, please see SI Section S7.

However, as prior research has also emphasized, increased contributions do not necessarily translate to improved collective outcomes (i.e., overall earnings) once punishment costs are taken into account (Dreber et al., 2008; Herrmann et al., 2008; Wu et al., 2009). To capture the effect of punishment on overall earnings, we examine *normalized efficiency*, which scales group earnings relative to the minimum in the absence of punishment (where no one contributes anything, "full defection") and the maximum in the absence of reward ("full cooperation") in each setting (see Materials and Methods for details). To illustrate, consider two PGGs that have the same configuration (number of players/rounds, information availability, etc.), with the exception that Game 1 has a multiplier of 2 and Game 2 has a multiplier of 4. A group that fully defects in Game 1 achieves a standard efficiency of 0.5, while a fully defecting group in Game 2 achieves a standard efficiency of 0.25, despite the two groups exhibiting the same degree of cooperation. Normalized efficiency addresses this discrepancy by scaling each group's earnings relative to the natural benchmarks of full cooperation and full defection respectively.

We find that although punishment consistently increased contribution levels, its effect on overall welfare was more complicated. As shown in Figure 2B, on average, punishment reduced normalized efficiency from 0.71 to 0.63 (an 11% decrease) in the learning experiments (coefficient = -0.089, SE = 0.032, 95% CI [-0.151, -0.027], p = 0.005, N = 335 groups), with a smaller but directionally consistent decrease from 0.72 to 0.68 (a 5.5% decrease) in the validation experiments (coefficient = -0.043, SE = 0.022, 95% CI [-0.087, 0.001], p = 0.056, N = 417 groups); for further details on regression, please see SI Section S7. These negative effects on efficiency align with several studies showing that punishment's costs can outweigh its benefits (Dreber et al., 2008; Herrmann et al., 2008; Wu et al., 2009). However, the overall average masks substantial and significant heterogeneity across PGG settings. As Figs. 2D and 2F show observed punishment effect sizes varied from strongly positive to strongly negative (in the learning experiments, we grouped conditions by their design parameters into 20 clusters using k-means for visualization and to enable standard error estimation; see Materials and Methods). Statistical tests of heterogeneity (Cochran's Q-statistic and Fisherian permutation tests; see Materials and Methods) consistently indicated that this variation exceeded what would be expected by chance, both in the learning [$Q(19) = 29.01$, $p = 0.066$; permutation tests $p < 0.01$] and validation experiments [$Q(19) = 30.29$, $p = 0.048$; permutation tests $p < 0.01$], as well as when aggregated across both waves [$Q(39) = 59.34$, $p = 0.019$; permutation tests $p < 0.01$], with between-experimental-condition differences explaining a substantial portion of this variation ($I^2 = 34\%$ in learning; $I^2 = 37\%$ in validation; see Materials and Methods).

Notably, because all experiments were conducted under identical protocols, recruitment procedures, and interfaces, this heterogeneity can only be attributed to differences in game parameters, unlike post-hoc meta-analyses where variation might stem from differences in research teams, time periods, populations, or experimental procedures (Holzmeister et al., 2024; Huedo-Medina et al., 2006). Therefore, our observed $I^2$ values represent a tighter measure of meaningful heterogeneity in the effect of punishment than is seen in typical meta-analyses. The magnitude of this heterogeneity is substantial. As shown in Figures 2C and 2D, our learning experiments suggested that punishment could dramatically reduce normalized efficiency from 0.78 to 0.44 (a 44% decrease) in some settings while increasing it from 0.56 to 0.80 (a 43% increase) in others, though these estimates relied on clustering to enable standard error estimation (see Materials and Methods). Most importantly, this pattern was replicated in our validation experiments, which provided a more reliable representation of heterogeneity through higher precision (8–12 trials per condition) and pre-registration, where punishment reduced efficiency from 0.72 to 0.40 (a 44% decrease) in some settings while increasing it from 0.63 to 0.81 (a 29% increase) in others (see Figure 2E and 2F).

The heterogeneity in punishment effects across experimental conditions presents a significant challenge for theory building (Bryan et al., 2021). Broad generalizations about punishment effectiveness based on individual PGG settings are inevitably brittle and may not hold even within seemingly similar cooperative contexts. Conversely, developing theories specific to only a single PGG configuration would be of very limited usefulness. A middle ground is to take an integrative approach that identifies the range of conditions over which punishment is likely to enhance or diminish cooperation efficiency. To this end, we frame the problem as an out-of-sample prediction exercise (Almaatouq et al., 2022; Hofman et al., 2017, 2021; Reichman et al., 2024; Rocca & Yarkoni, 2021; Yarkoni, 2020; Yarkoni & Westfall, 2017), where results acquired from one set of experimental conditions can be used to predict punishment effectiveness in entirely new settings. The associated models can, in turn, drive theory development by illuminating important relationships between game parameters and punishment effectiveness.

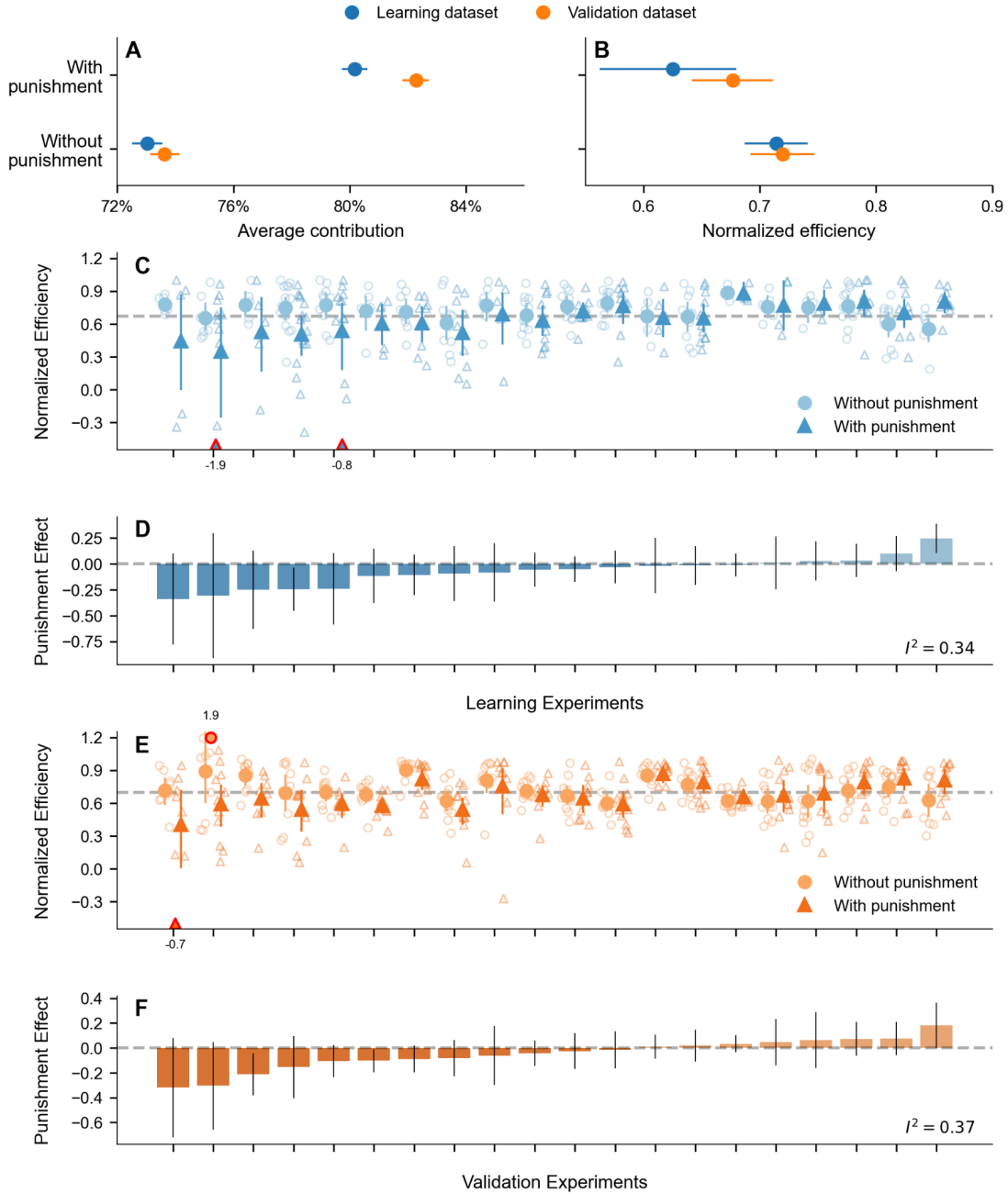

**Figure 2. Average punishment effects and heterogeneity across conditions.** Panel (A) shows the average contribution per round (as a percentage of per round endowment) to the public good in games with and without punishment. Panel (B) shows normalized efficiency in games with and without punishment. For the learning experiments, Panel (C) shows normalized efficiency across different experimental conditions with and without punishment (filled markers show means, open markers show individual games) and Panel (D) shows the resulting punishment effects on normalized efficiency. Panels (E) and (F) show the corresponding analyses for validation experiments, where multiple trials per condition enabled direct estimation without clustering. The dashed line in Panels (C) and (E) represents the mean normalized efficiency in the learning and validation experiments respectively. Points outside the y-axis range are shown at the boundary with red outlines and labeled with their actual values. In the learning experiments, we grouped conditions by their design parameters into 20 clusters to enable standard error estimation. Error bars show 95% confidence intervals. See SI Section S8 for clustering details.

**Predicting punishment effectiveness in new experiments.** To operationalize our out-of-sample prediction approach, we leveraged our two-wave data collection process. Specifically, using PGG parameters and the efficiency level in the control condition (punishment disabled), we fit several statistical models—elastic net (E-Net) with two-way interactions, ordinary least squares (OLS), random forest (RF), multilayer perceptron (MLP), and XGBoost (XGB)—to predict efficiency in the treatment condition (punishment enabled) in our Wave 1 learning experiments. We then used these fitted models to predict efficiency in the Wave 2 experiments run using 20 new sets of game parameters that were held out for validation. This prediction task reflects the scenario in which a decision-maker observes the status quo (the control condition, in which punishment is disabled) and could use the prediction of the outcome under treatment to decide whether to introduce peer punishment. Because in this scenario the decision-maker is only ever predicting the effect of punishment for a single game configuration, we use our pre-registered outcome of regular efficiency (net earnings scaled by the full cooperation baseline; see Materials and Methods for details) rather than normalized efficiency on the grounds that the former is more intuitive to human judges, whose predictions we will compare with our models.

We assessed predictive performance using both root mean squared error (RMSE) and out-of-sample $R^2$. The latter measures how much better our model predicts relative to simply using the average treatment efficiency (i.e., the efficiency of the PGG conditions with punishment available) from the learning experiments (see the Model Evaluation section in Materials and Methods). Effectively, $R^2$ quantifies the proportion of out-of-sample variance explained by our model, or a human judge, compared to this baseline; that is, an $R^2$ of 0 indicates predictions no more accurate than always predicting the mean treatment efficiency observed in the learning experiments. Correspondingly, $R^2$ is bounded above by 1 (perfect prediction) but has no lower bound as predictions can be arbitrarily worse than the baseline. To contextualize the predictive power of our models, we invited domain experts (53 academics: six authors of peer-reviewed PGG papers and 47 forecasters from the Social Science Prediction Platform) and laypeople (500 Prolific workers) to complete the same prediction task. From these individual-level predictions, we derived a collective "wisdom of the crowd" benchmark within each group, defined as the mean prediction across all forecasters.

As shown in Figures 3A–D, the E-Net model achieved the best predictive performance ($R^2$ = 0.53, RMSE = 4.52, Fig. 3A), followed by OLS ($R^2$ = 0.38, RMSE = 5.22, Fig. 3B). The collective predictions of both domain experts and laypeople performed similarly to baseline ($R^2$ = 0.02, RMSE = 6.53, Fig. 3C; $R^2$ = 0.05, RMSE = 6.44, Fig. 3D, respectively). Strikingly, the superiority of the model over humans applied not only to their averaged predictions but to individuals as well: not one out of 553 human judges outperformed the model. This performance gap reveals that human judges struggle to mentally integrate how multiple factors combine to determine outcomes. Notably, experts performed no better than laypeople, suggesting that identifying which factors matter (the focus of most prior work) does not translate into understanding how they combine to produce outcomes.

While interpreting specific numerical values of RMSE and R² as "good" or "bad" in an absolute sense remains challenging, the E-Net model clearly outperforms both a "null" model that always predicts the average treatment effect from training data as well as the judgments of laypeople and experts. The E-Net's out-of-sample R² of 0.53 also compares favorably to recent prediction exercises in the social sciences, such as predicting individual life-course outcomes (Rigobon et al., 2019; Salganik et al., 2020), Twitter cascade sizes (Martin et al., 2016), and group performance on novel instances of a single task (Almaatouq, Alsobay, et al., 2024), for which R² varies between 0.05 and 0.4.

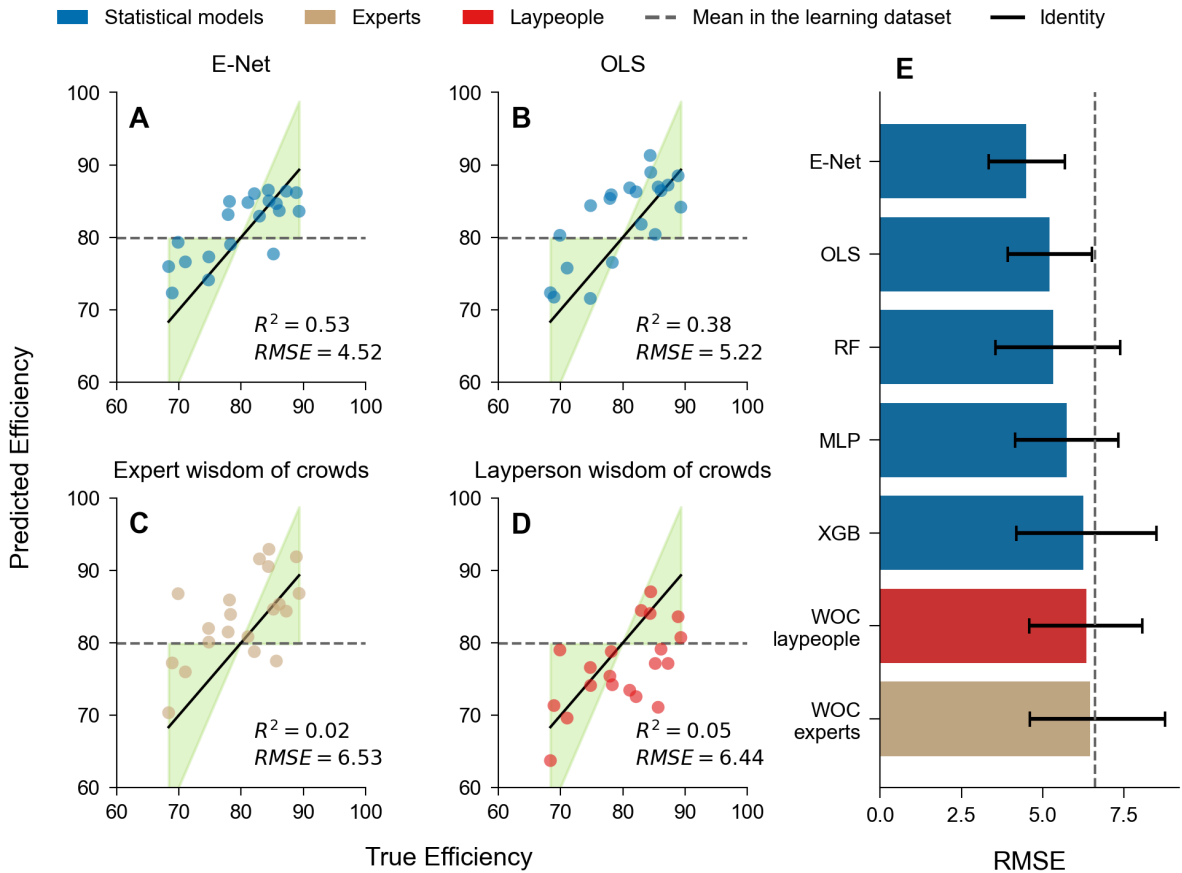

**Figure 3. Predictive performance on unseen experiments.** Panels (A–D) show the predicted versus true efficiency values for the 20 validation experiments. The diagonal line represents perfect prediction. The dashed red line shows the baseline prediction, while the regions shaded in green indicate where predictions have lower absolute error than the baseline. Each panel includes the out-of-sample R² and root mean squared error (RMSE) of the predictions. All predictions were made using the PGG design parameters and efficiency in the control condition (punishment disabled) to predict efficiency in the treatment condition (punishment enabled). The observed performance metrics in Panels (A-D) are descriptive quantities for our specific validation set, but one can also view these experiments as a random sample from the population of all possible PGG configurations. Panel (E) shows bootstrapped 95% confidence intervals for RMSE across prediction approaches. The E-Net model outperformed both experts (95% CI of difference in RMSE: [-0.10, 3.71]) and laypeople (95% CI of difference in RMSE: [0.03, 3.86]), while experts and laypeople showed no significant difference in performance (95% CI of difference in RMSE: [-2.4, 2.6]). See SI Section S9 for statistical model details, calculation of out-of-sample R², confidence interval construction, and additional predictive baselines.

**Key PGG design dimensions for predicting punishment effectiveness.** Beyond simply being able to predict outcomes, our models also help contribute to theory building by shedding light on the specific conditions under which punishment is helpful versus harmful for welfare. Specifically, we assess which features (i.e., parameters) of our best-performing model (E-Net) are most important for predicting punishment effectiveness, using two complementary approaches (analyses for other statistical models are presented in SI Section S11). First, we use permutation analysis (Breiman, 2001) to examine each feature's importance for prediction accuracy by randomly shuffling the values of PGG parameters in the validation experiments and measuring the resulting percentage change in prediction error (Fig 4A). Features that cause larger increases in prediction error when shuffled are more important for prediction. Second, we use SHAP (Shapley Additive Explanations) values (Lundberg & Lee, 2017), which quantify how each feature contributes to the model's individual predictions rather than its overall performance (Fig 4B). These measures are complementary, as permutation feature importance reveals which features are most predictive of punishment effectiveness, while SHAP values show how these features influence the model's predictions. Both are important, as overall feature importance does not necessarily imply consistency of effect or vice versa. Using this dual ranking system, we next describe features that fall into one of three categories: features such as peer communication that are important and have consistently positive or negative effects; features such as contribution framing or game length that are important but whose effects differ in direction (or magnitude) depending on interactions with one or more other features; and features such as MPCR that show consistent directionality of effects, but low predictive importance (we do not describe features that are neither predictively important nor consistent in their directionality).

*The important (and consistent) role of communication.* Although communication has long been recognized as an effective mechanism for promoting cooperation in social dilemmas (Balliet, 2010; Bochet et al., 2006; Bouas & Komorita, 1996; Ostrom et al., 1994; Ostrom & Walker, 1991), our results show it is also the dominant feature in determining punishment effectiveness. Permutation feature importance shows that shuffling this feature increased prediction error by 60%—more than three times that of the next most important feature (Fig 4A). SHAP values further show that allowing communication has an unambiguously positive impact on punishment effectiveness (Fig 4B). This quantification of relative importance—communication being three times more influential than any other factor—represents a different type of knowledge than simply knowing communication "helps" and is precisely what single-factor studies cannot reveal. Given that punishment itself is a public good (i.e., everyone benefits from increased cooperation when free-riders are deterred, but no individual is incentivized to bear the cost of punishing), communication might reinforce social norms by allowing groups to explicitly establish expectations about both cooperation and punishment, potentially reducing arbitrary or retaliatory punishment while ensuring consistent sanctions for non-cooperation (Andrighetto et al., 2013; Putterman, 2010). It could also enable groups to coordinate their sanctioning efforts, preventing both over-punishment of some free-riders and under-punishment of others (Boyd et al., 2010; Molleman et al., 2019).

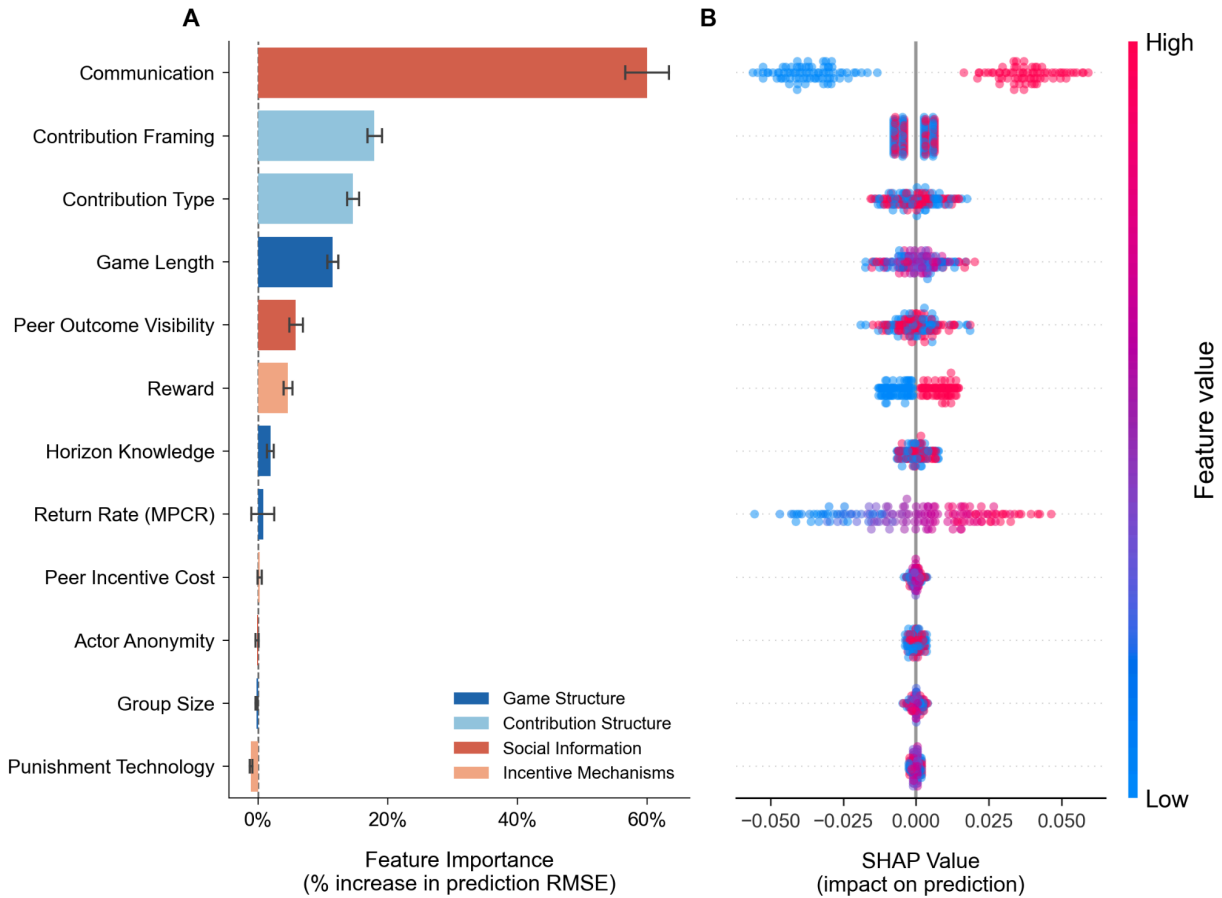

**Figure 4. Feature importance and model interpretation.** Panel (A) shows each feature's importance to out-of-sample prediction. For each PGG design parameter, the parameter's importance to the E-Net model's predictive performance is measured as the ratio of the model's error in predicting the outcome of validation experiments when the values for that parameter are randomly shuffled, and its baseline error on the original, unpermuted set of validation experiments. Each parameter is shuffled independently, and error bars indicate 95% confidence intervals arising from the random parameter shuffles. Panel (B) displays SHAP values showing how each feature contributes to individual predictions. Each point represents one learning experiment, with color indicating feature value (e.g., for communication: blue = disabled, pink = enabled) and horizontal position showing the magnitude and direction of the contribution to predicted efficiency. Features whose points appear primarily on the right increase predicted efficiency, while those on the left decrease it. The horizontal spread reflects how a feature's impact varies across experiments. For instance, while communication generally increases efficiency when enabled (pink dots mostly on the right), its magnitude depends on other parameters in each specific game configuration. Note: The model also uses the control (no punishment) efficiency to inform predictions, although it is omitted from the figure for clarity, as it was not a design space parameter. Similar analyses for the other statistical models are presented in SI Section S11.

*Features that consistently enhance punishment (but vary in importance).* Two other features showed consistently positive effects on punishment effectiveness, regardless of other game parameters. First, the availability of reward mechanisms increased efficiency in the punishment conditions (Fig 4B) and was important for prediction accuracy (4.6% increase in error when shuffled; Fig 4A). Although the relative effectiveness of rewards versus punishment has long been a matter of debate in the literature (Andreoni et al., 2003; Balliet et al., 2011; Baron, 2009; Bravo & Squazzoni, 2013; Gao et al., 2015; Rand et al., 2009), our results show that the availability of rewards consistently enhanced punishment's effectiveness. Second, higher marginal per capita return (MPCR), which reduces the conflict between individual and collective interests, enhanced punishment effectiveness, though its effect was too small relative to other features to meaningfully impact prediction accuracy.

*Features that are important but have contingent effects.* Several other features, such as contribution framing (opt-out vs. opt-in), contribution type (variable vs. all-or-nothing), game length, and visibility of peer outcomes, rank highly in importance but exhibit effects that are contingent on other features. To illustrate, we focus on two of these features: game length and contribution framing.

First, game length substantially influences prediction accuracy (11.5% increase in error when shuffled), but its effect on punishment effectiveness depends on two other features: the availability of communication and the visibility of peer outcomes. As shown in Fig 5A and 5B, longer games enhance punishment effectiveness only when communication is available, and this positive interaction is weaker when peer outcomes are visible. These interactions suggest that punishment becomes more effective with communication over repeated iterations of the game because groups may establish and maintain cooperative norms that, once established, reduce the need for frequent punishment (Andrighetto et al., 2013; Gächter et al., 2008; Putterman, 2010). The visibility of peer outcomes seems to dampen this positive effect. Although our experiment cannot identify the mechanism, visibility might reduce punishment effectiveness in multiple ways: participants who observe non-contributors or non-punishers earning higher payoffs may lose motivation to enforce cooperative norms, while the ability to identify and retaliate against punishers could make individuals more hesitant to sanction free-riders (Dreber et al., 2008).

Second, we find that the contribution framing (opt-out vs. opt-in) strongly influences prediction accuracy (18% increase in prediction error when shuffled). In standard public goods games (opt-in), participants start with their endowment in their private account and must actively choose to contribute to the public good. Under the opt-out framing, this default is reversed: endowments begin in the public good, and participants must actively choose to withdraw funds for private use. Importantly, individuals can only withdraw up to their total endowment, ensuring that both opt-in and opt-out framings are monetarily equivalent. Previous research has shown mixed effects of framing on cooperation (Andreoni, 1995; Brewer & Kramer, 1986; Cox & Stoddard, 2015; Fleishman, 1988). Our results reveal a contingent relationship between framing and punishment effectiveness, showing that the impact of opt-out framing depends primarily on whether contributions must be all-or-nothing or can be variable/partial (Fig 5C). The opt-out default enhances punishment effectiveness when participants can make variable contributions but, surprisingly, reduces effectiveness when contributions are restricted to all-or-nothing. This interaction is further modulated by peer outcome visibility: having outcomes be visible amplifies the negative effect of opt-out framing with all-or-nothing contributions but attenuates the positive effect with variable contributions (Fig 5D).

*Punishment parameters matter less than expected.* In addition to these complex and non-obvious positive effects, our approach also reveals some surprising null results. For example, Fig 4A shows that punishment technology, which is defined as the impact of a punishment action relative to its cost and hence seems a likely candidate to impact efficiency, had the smallest overall effect on the model's predictive performance of all the features we tested, scarcely different from zero. This unexpected finding could be interpreted in at least two ways. One interpretation suggests that punishment's mechanical design may not be a determinant of its success and that its effectiveness depends more on the context in which it operates—factors like communication availability, contribution framing, and interaction dynamics. Another explanation reflects methodological limitations such as model overfitting, where our model may be capturing idiosyncrasies in the learning data regarding the punishment technology rather than generalizable patterns.

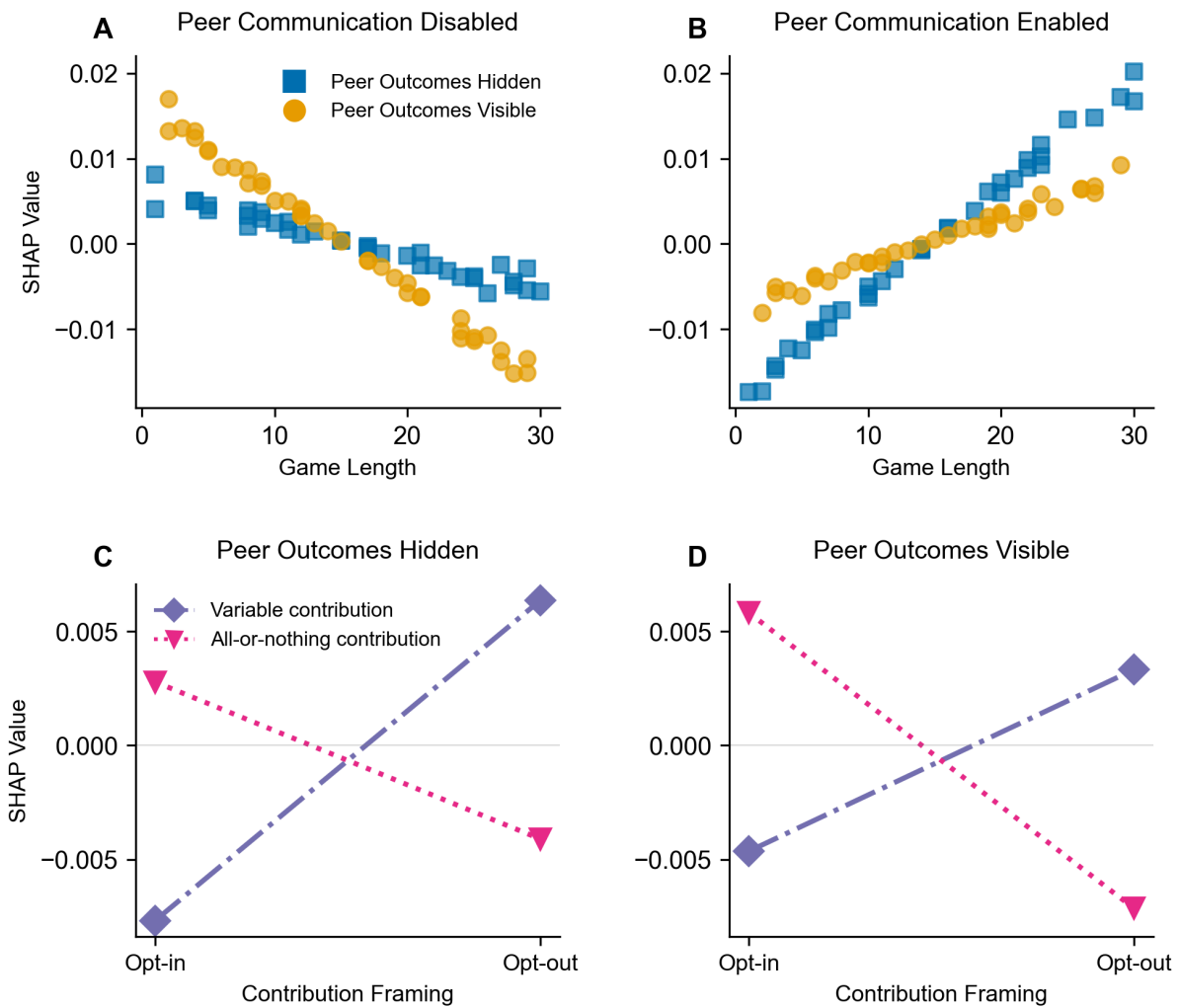

**Figure 5. Key interaction effects shaping punishment effectiveness.** Panels (A) and (B) show how game length interacts with communication and peer outcome visibility. Longer interactions enhance punishment effectiveness only when communication is available, and this positive effect is weaker when peer outcomes are visible. Panels (C) and (D) illustrate the interaction between contribution framing and contribution type. The opt-out default enhances punishment effectiveness with variable contributions but reduces it with all-or-nothing contributions. The effect is amplified for all-or-nothing contributions and attenuated for variable contributions when peer outcomes are visible.

## Discussion

Returning to our motivating question (what is the effect of punishment on cooperation in public goods games?) our integrative experiment reveals that the answer depends critically on configurations of factors that traditional approaches cannot capture. The dramatic heterogeneity in punishment effects (from improving efficiency by 43% to reducing it by 44%) reflects systematic patterns rather than noise, as demonstrated by our model's ability to predict outcomes in entirely new experimental conditions, which demonstrates that integrative experiments can extract these systematic patterns, transforming qualitative insights into quantitative predictions about when punishment will help or harm cooperation.

Three key patterns emerge. First, communication dominates all other factors by a factor of three (a quantitative insight about relative importance that single-factor studies cannot reveal). This quantification of relative importance represents a different type of knowledge than simply knowing communication "helps," and is precisely what single-factor studies are not designed to reveal. Second, crucial interaction effects explain previous contradictory findings: game length enhances punishment only with communication present, while contribution framing effects depend on both contribution flexibility and outcome visibility. These interactions are inherently invisible when holding "all else constant." Third, the most-studied parameters (i.e., punishment cost, magnitude, and impact) have negligible effects, while rarely-examined factors like contribution defaults prove critical. This mismatch between research attention and actual importance suggests the field may have been allocating resources studying the mechanisms behind secondary parameters while overlooking primary determinants of punishment effectiveness.

These findings demonstrate why integrative experiments combined with machine learning offer a powerful approach for studying complex social phenomena. Our analyses (permutation feature importance identifying which features matter most and SHAP values illuminating how they influence outcomes) allow us to navigate the tension between predictive accuracy and interpretability. While the three-way interactions are challenging to interpret and often counterintuitive, they offer a more accurate characterization of when punishment enhances or undermines cooperation than existing theoretical frameworks and conventional explanatory narratives.

Several aspects of punishment effectiveness warrant further investigation. Our study focused exclusively on design parameters of public goods games, holding population characteristics constant. Previous research has demonstrated substantial variation in punishment effectiveness across populations (Balliet & Van Lange, 2013; Henrich et al., 2006; Herrmann et al., 2008). Our results complement this work by showing variation across different PGG contexts within the same population. Although recent evidence suggests that experimental design characteristics may have a larger effect than population characteristics (Holzmeister et al., 2024), the interaction between population and context variation remains largely unexplored. That we find such substantial heterogeneity even within the most well-studied population group suggests that studying these interactions could reveal even more complex patterns, and therefore that our prediction exercise likely represents an optimistic assessment of performance, both for quantitative models and experts.

Similarly, while our 14-dimensional design space captures many key aspects of cooperative settings, it is not intended to be comprehensive or definitive. Rather, our contribution should be viewed as the first step in an iterative process of design space construction and empirical testing (Almaatouq, Griffiths, et al., 2024). Design parameters serve as features whose utility can be empirically evaluated through out-of-sample prediction. Those that fail to predict punishment effectiveness may be removed or refined, while predictively important features will be retained and potentially disaggregated into more fine-grained constituent features (e.g., "communication" could be further decomposed into its semantic or structural components). In this way, unexpected findings can guide the modification of existing dimensions or suggest new ones, allowing the framework to evolve based on clear empirical criteria (i.e., its ability to predict punishment effectiveness in new settings) rather than relying solely on theoretical arguments or statistical relationships in existing data.

Finally, the dataset produced by our integrative experiment is itself a contribution to the development of machine learning models in behavioral science (Almaatouq et al., 2022; Peterson et al., 2021; Reichman et al., 2024; Zhu et al., 2024). Unlike behavioral datasets that emerge from testing specific hypotheses, our data was generated through space-filling and random parameter space exploration, making it particularly valuable for training and validating predictive models. Just as large, purpose-built datasets have accelerated progress in other fields, we anticipate that this dataset will facilitate the development of more sophisticated models for predicting human behavior in social dilemmas. Datasets of this sort could be particularly valuable as researchers increasingly employ machine learning to understand complex social phenomena.

We conclude by noting that the approach we have implemented here represents a promising path forward for social science research. Rather than debating whether punishment "works," our approach addresses the "solution oriented" (Watts, 2017) question of under what conditions it does and does not work, which is a question that can only be answered through systematic investigation of the relevant parameter space. In addition to shedding light on the focal question of punishment in public goods games, this approach can be extended to many other domains in social science in which the context sensitivity (Goroff et al., 2018; Manzi, 2012) of the phenomena in question has produced mixed and seemingly contradictory findings. Integrative experimental design, combined with modeling that integrates prediction and explanation (Hofman et al., 2021), provides a more reliable and comprehensive characterization of complex social behaviors.

# Materials and Methods

**Participant Recruitment**

A total of 7,100 participants were recruited from a curated panel on Amazon Mechanical Turk (N=604) and Prolific (N=6,496) and randomly assigned to groups to play synchronous PGGs of varying design. Participants were monetarily incentivized to maximize their individual in-game earnings. Base pay, per round pay, and the conversion rate from in-game currency to USD were constant across all studies; other parameters that influence pay, such as the marginal per capita return, number of rounds, group size, and the cost and impact of punishment/reward, varied across games. Throughout the study, participants were only allowed to participate once.

The study was determined to be exempt under Category 3 (Benign Behavioral Intervention) by the Committee on the Use of Humans as Experimental Subjects at the Massachusetts Institute of Technology (Protocol #E-5462). All participants provided explicit consent. For details on recruitment materials and the incentive structure, see SI Section S5.

**Experiment Implementation**

The experiment was developed using the Empirica platform (Almaatouq et al., 2021). For screenshots of the interface and its implementation details, please refer to SI Section S6.

Our PGG is implemented as a series of rounds, where each round is composed of three consecutive stages: (1) In the *contribution* stage, players are each granted a per round endowment and decide how much of the endowment to contribute to the public fund. (2) In the *redistribution* stage, players are shown the coins contributed by each player, as well as the sum of all contributions, and the amount redistributed to each player. During this stage, players may punish and/or reward other players, as applicable to the configuration of the PGG. (3) In the *round summary* stage, players view a summary of their earnings, including their retained endowment, share of the public fund payoff, and punishments/rewards that were given/received; depending on the PGG configuration, they may also view this information for other players. No actions are taken during this stage. After the *round summary* stage is complete, another round begins starting from the *contribution* stage.

Before entering a game, players complete an interactive walkthrough of the actual interface used in-game, with a question to check their comprehension after each stage's explanation. To ensure active participation throughout the duration of the games, we also implement measures to detect and remove idle players—for details of the walkthrough and idle detection, please see SI Sections S5 and S6. As detailed in Table 1, 14 PGG parameters were actively manipulated throughout the study. Other factors were held constant across all PGG configurations, including the duration of all stages (45 seconds), the per round endowment (20 coins), and the details of the financial incentive structure (base pay, per round participation pay, lobby pay, and the conversion rate from coins to USD). For details on parameter definitions and their implications on the game's mechanics and interface, please refer to SI Section S6.

**Study Design**

The experiments in this study were executed in two waves: (1) A "learning" wave in which experiments were selected with the goal of training models to predict experimental outcomes, and (2) a "validation" wave to evaluate the out-of-sample predictive performance of models trained using the data from the learning phase.

In the learning experiments (Wave 1), 3,618 participants (604 from Amazon Mechanical Turk, 3,014 from Prolific) were recruited to play 335 PGGs of varying configurations. The configurations in the learning experiments were determined by a Sobol sequence (Joe & Kuo, 2008; Owen, 1998; Virtanen et al., 2020) with the goal of achieving more uniform coverage of the PGG design space than expected by random sampling. Each experiment in this wave was executed once, following the rationale that, all else being equal, when studying heterogeneity, precision and power are greater when sampling many conditions with fewer observations each (Kenny & Judd, 2019; Moerbeek & Teerenstra, 2021; Vandekerckhove, 2024). Our approach here aligns with methodological innovations that explicitly design for heterogeneity across experimental contexts (Awad et al., 2018; Baribault et al., 2018; DeKay et al., 2022; Kenny & Judd, 2019; Moerbeek & Teerenstra, 2021; Peterson et al., 2021; Zhu et al., 2024). To facilitate standard error estimation and the visualization of punishment effect heterogeneity across learning experiments in Figure 2C, the configurations in this wave were divided into 20 clusters by k-means clustering with min-max scaling for integer-valued parameters (e.g., game length, group size). For implementation details, including the generation and use of the Sobol sequence to select PGG conditions and alternative clustering approaches, please see SI Sections S4 and S8.

In the validation experiments (Wave 2), 3,482 participants (all from Prolific) were recruited to play 417 PGGs of 40 different configurations. The configurations in the validation wave were chosen by randomly assigning each parameter value independently and ensuring that the resulting configurations were not also used in the learning experiments.

The numbers of configurations, sampled games, and players reported are counted after filtering data from games affected by server failures and insufficient participant recruitment. In the learning wave, 6 games with a multiplier of 1 (for which normalized efficiency is undefined) and a single game in which participants cooperated to establish a system of mutual reward were also filtered. For complete details on configuration selection and data collection/filtration, please see SI Sections S2 and S4.

**Pre-registration Details**
The randomly selected configurations in the validation wave, illustrated in Figures 2E and 2F, were pre-registered ([AsPredicted #182022](AsPredicted #182022)) to declare the selected conditions and sample size before running the validation experiments. Additionally, the details of the human forecasting task used to benchmark model performance (including the predictive target, model predictions, comparisons of human and model predictions, and recruitment strategy), illustrated in Figure 3, were pre-registered ([AsPredicted #191059](AsPredicted #191059)) before collecting the human forecasts.

All other analyses in this work are exploratory. For deviations from the pre-registration and associated sensitivity analyses, please see SI Section S1.

**Calculating Cooperation Efficiency**
In PGG literature, a commonly measured outcome is the "efficiency" of a group, defined as the ratio of $E_{group}$, the group's total earnings (net of all costs and penalties/rewards) to $E_{full\ cooperation}$, the earnings of a hypothetical, fully cooperative group in which every member contributes their full endowment in every round and never punishes or rewards. Mathematically, this is expressed as:

$$Efficiency = \frac{E_{group}}{E_{full\ cooperation}}$$

In our study, this conventional efficiency measure is only used in our human forecasting task, as it is familiar to expert forecasters, and appropriate for comparing treatment and control conditions with the same PGG configuration.

However, to compare cooperative behavior across different PGG configurations, we introduce a measure of "normalized efficiency," which addresses differences in the range of possible efficiencies across conditions. For instance, in a PGG where the fund multiplier is 1x and both punishment and reward are disabled, a group's efficiency would always be 1, even if all members defect. By scaling the group's earnings relative to the minimum and maximum possible earnings in each condition, we facilitate direct comparison between cooperative outcomes in PGGs of varying designs. Mathematically, normalized efficiency is expressed as:

$$Normalized\ Efficiency = \frac{E_{group} - E_{full\ defection}}{E_{full\ cooperation} - E_{full\ defection}}$$

where $E_{full\ defection}$ is the total earnings of a hypothetical group in which all members fully defect (i.e., contribute nothing to the public fund) in every round and never punish or reward. We note that normalized efficiency can exceed 1 when rewards are enabled (e.g., in a setting with full cooperation when participants also reward one another) and can be negative when punishment is enabled (e.g., zero cooperation with punishments). Nonetheless, normalized efficiency acts as a useful, unitless measure that aids comparisons across different game configurations.

To account for potential participant dropout in games, all earnings (both real and hypothetical) are calculated at the round level and then aggregated at the game level. For example, if a PGG began with N players, 1 of whom dropped out in round R, the hypothetical earnings of a fully cooperative/defective group playing this game would assume N players up to round R-1, and N-1 players in rounds including and following round R.

**Heterogeneity Tests**

To detect and measure between-design heterogeneity in the effect of punishment on cooperation, we employ three complementary approaches. While Cochran's Q and Fisherian randomization testing allow us to test for the existence of *any* heterogeneity in the underlying effect, the complementary $I^2$ measure allows us to compare the degree of underlying variation relative to variation expected by random sampling error. In all three approaches, outcomes are measured using normalized efficiency. Because the scale of normalized efficiency is the same across experimental settings, effects measured in normalized efficiency can be compared directly without further standardization. Where relevant, all statistical tests in this study are two-sided.

(1) Cochran's Q-statistic (Cochran, 1954; Higgins et al., 2003; Huedo-Medina et al., 2006) allows us to test the null hypothesis that observed differences in effects are due solely to chance, rather than real variation in the true underlying effects. Mathematically, Cochran's Q represents the total squared deviation of point estimates from the meta-analytic estimate (the average of point estimates weighted inversely by their variance) and follows a chi-squared distribution with $n - 1$ degrees of freedom, where $n$ is the number of estimates:

$$w_i = 1/\hat{\sigma}_i^2 \quad , \quad \bar{T} = \frac{\sum_i w_i T_i}{\sum_i w_i}$$

$$Q = \sum_i w_i (T_i - \bar{T})^2$$

where $T_i$ is the i'th point estimate, $\hat{\sigma}_i^2$ is the estimate of its variance, and $w_i$ is its weight under a fixed-effects meta-analytic model. Point estimates are calculated within each cluster in the learning wave and within each experiment in the validation wave.

(2) Instead of relying on point estimates made within clusters/experiments, as in Cochran's Q, the Fisherian randomization test (FRT) relies on the shifted Kolmogorov-Smirnov (K-S) statistic (Ding et al., 2016) between two distributions: trial-level outcomes in the control (without punishment) group shifted by a given constant treatment effect, and trial-level outcomes in the treatment (with punishment) group. Since the true treatment effect is unknown, the procedure tests 300 treatment effect values from a 99.9% confidence interval around the estimated treatment effect. For each treatment effect value, the treatment assignments are randomized 1000 times; for each randomization, the shifted K-S statistic is calculated, forming the null distribution of the statistic for the current treatment effect value from which the p-value of the observed K-S statistic is calculated. The procedure is summarized by the maximum p-value found across all treatment effect values used.

(3) While Cochran's Q and the FRT provide omnibus tests for the existence of heterogeneity based on treatment effect estimates and outcome distributions respectively, they do not describe the degree of heterogeneity that exists. Complementing these tests, $I^2$ estimates the proportion of variation in point estimates that is due to variation in the underlying effect as opposed to within-estimate sampling error (Higgins et al., 2003; Huedo-Medina et al., 2006), and is defined as:

$$I^2 = \frac{Q - (n - 1)}{Q}$$

**Model Construction**

We defined the prediction task in the study as predicting the average group efficiency when punishment is enabled, given the design parameters of a PGG and the average group efficiency when punishment is disabled.

Five statistical models were trained to complete this prediction task using the data from the learning experiments: ordinary least squares regression (OLS), elastic net regularized regression (E-Net) (Zou & Hastie, 2005), a random forest (RF) (Breiman, 2001), XGBoost (XGB) (Chen & Guestrin, 2016), and a multilayer perceptron (MLP, also known as a dense neural network). All models used the raw features, except for the E-Net, which used standardized features and their pairwise interactions, and the MLP, which used standardized features. For the E-Net, RF, XGB, and MLP models, hyperparameters were selected through a Bayesian optimization process maximizing cross-validated predictive performance on data from the learning experiments. For details including data pre-processing, hyperparameter ranges, optimization implementation, and optimal configurations, please refer to SI Section S9.

**Model Evaluation**

Models were primarily evaluated by measuring the root mean squared error (RMSE) of their predictions of average group efficiency under punishment in the validation wave, the data for which was held out during model training. A secondary measure of performance is out-of-sample $R^2$ (referred to simply as $R^2$), defined as:

$$R^2 = 1 - \frac{\sum_i (y_i - \hat{y}_i)^2}{\sum_i (y_i - \bar{y}_{learn})^2}$$

where $i$ denotes the index of the experiment in the validation wave and $\bar{y}_{learn}$ is the average group efficiency under punishment in the learning experiments. For evaluation details, including the performance of various statistical benchmarks, please see SI Section S9.

**Human Forecast Elicitation**

To contextualize the predictive performance of the 5 statistical models, we compare their predictions to those of 53 domain experts recruited through the Social Science Prediction Platform (SSPP) (DellaVigna et al., 2019)—47 from the SSPP's general pool of social scientists and 6 from direct outreach to authors of papers studying PGGs—and 500 laypeople recruited through Prolific. Forecasters were informed that they would be judged by the squared error of their predictions; while SSPP forecasters were uncompensated, Prolific workers were incentivized with a base pay of $2.00 and a maximum bonus of $0.50 for each prediction based on prediction accuracy (for a total of $10.00 in potential performance bonus). For both experts and laypeople, we define the "wisdom of crowds" prediction of average treatment efficiency to be the mean prediction made by the group. For details on expert and layperson recruitment and compensation, the prediction survey, and analyses of individual predictor performance, please see SI Section S10.

**Model Interpretation Methods**

To gauge the importance of each individual feature to our models' behavior, we report two complementary measures: permutation feature importance (PFI) (Breiman, 2001) and Shapley additive explanations (SHAP) (Lundberg & Lee, 2017). While PFI measures importance to out-of-sample predictive performance, SHAP values measure the marginal impact, and its direction, of a feature's value on the prediction made by a model.

PFI is calculated by first establishing a model's baseline RMSE on the unpermuted validation set. Then, for each feature being evaluated, its values are randomly shuffled in the validation set while keeping other features unchanged; the model's RMSE on the shuffled data is then measured, and the degradation in performance (relative to the baseline) indicates that feature's importance. This process is repeated 30 times to quantify the uncertainty over shuffles of a feature's value.

For implementation details and PFI/SHAP calculations across all statistical models, please see SI Section S11.

For discussions of the use of machine learning in predictive and integrative approaches in social science and their relationship to traditional methods, we refer to Hofman et al. (2021), Yarkoni & Westfall (2017), and Molnar & Freiesleben (2024).

# Data Availability

The datasets generated and analysed during the current study are available in the Open Science Framework (OSF) repository: https://osf.io/2d56w/?view_only=d046c1c417024569a8f9fed9e6c8d4d1

# Code Availability

The code to conduct the analyses in the current study and its supplementary materials, as well as the code for the PGG experiment platform, are available in the Open Science Framework (OSF) repository: https://osf.io/2d56w/?view_only=d046c1c417024569a8f9fed9e6c8d4d1

# Supplemental Information for
# Integrative Experiments Identify How Punishment Impacts Welfare in Public Goods Games

Mohammed Alsobay[a], David G. Rand[a], Duncan J. Watts[b,c,d], and Abdullah Almaatouq[a,e,*]

[a] Sloan School of Management, Massachusetts Institute of Technology
[b] The Wharton School, University of Pennsylvania
[c] Department of Computer and Information Science, University of Pennsylvania
[d] Annenberg School of Communication, University of Pennsylvania
[e] Institute for Data, Systems, and Society, Schwarzman College of Computing, Massachusetts Institute of Technology

* Correspondence to: amaatouq@mit.edu

# S1. Pre-registration, deviations, and code/data access

For access to project materials, including raw data and processing/analysis code, the code used to select PGG configurations in the learning and validation waves, and the code and documentation for the experimental platform, please visit the study's Open Science Foundation (OSF) repository (https://osf.io/2d56w/?view_only=d046c1c417024569a8f9fed9e6c8d4d1).

Two aspects of the study were pre-registered: (1) the PGG configurations randomly selected for the validation wave of experiments, and (2) the comparison of statistical model predictions to those of expert and lay forecasters.

The conditions for the validation wave were pre-registered in [AsPredicted #182022](AsPredicted #182022). We note the following errata and deviations:

- For configurations where reward is disabled, reward cost and magnitude should be displayed as "N/A" in the pre-registration; the values shown for those configurations are a typo.
- In our pre-registration, we describe the exclusion criterion as "if the total number of players who drop out is greater than 18% of the intended total number of players, the entire game will be discarded." After we collected the data and implemented this criterion, we realized this was an ambiguous description of our intent. As implemented in the study, we discard a game if the total number of players who drop out *of the game lobby (i.e., pre-game)* is greater than 18% of the intended group size.
An alternative interpretation of the criterion is to discard games where the total number of players who drop out of the game *at any time* is greater than 18% of the intended group size. The impact of this alternative interpretation is minimal and fully described in Section S3.
- During collection of the validation set, administrative and technical issues led to deviations from the target of collecting 12 experimental trials per configuration. These deviations are described in Table S1.

**Table S1. Intended, collected, and retained trials by configuration in the validation wave**

| | Configuration | Intended trials | Collected | Valid trials |
|---|---|---|---|---|
| 1 | Control | 12 | 12 | 10 |
| 1 | Treatment | 12 | 12 | 12 |
| 2 | Control | 12 | 12 | 8 |
| 2 | Treatment | 12 | 12 | 8 |
| 3 | Control | 12 | 12 | 9 |
| 3 | Treatment | 12 | 12 | 11 |
| 4 | Control | 12 | 10 | 9 |
| 4 | Treatment | 12 | 11 | 10 |
| 5 | Control | 12 | 12 | 12 |
| 5 | Treatment | 12 | 10 | 9 |
| 6 | Control | 12 | 12 | 9 |
| 6 | Treatment | 12 | 12 | 12 |
| 7 | Control | 12 | 12 | 12 |
| 7 | Treatment | 12 | 12 | 11 |
| 8 | Control | 12 | 12 | 11 |

| 8  | Treatment | 12 | 12 | 12 |
| 9  | Control   | 12 | 12 | 12 |
| 9  | Treatment | 12 | 12 | 11 |
| 10 | Control   | 12 | 12 | 11 |
| 10 | Treatment | 12 | 12 | 11 |
| 11 | Control   | 12 | 11 | 10 |
| 11 | Treatment | 12 | 11 | 10 |
| 12 | Control   | 12 | 11 | 10 |
| 12 | Treatment | 12 | 11 | 11 |
| 13 | Control   | 12 | 12 | 12 |
| 13 | Treatment | 12 | 12 | 12 |
| 14 | Control   | 12 | 11 | 10 |
| 14 | Treatment | 12 | 11 | 11 |
| 15 | Control   | 12 | 12 | 12 |
| 15 | Treatment | 12 | 12 | 10 |
| 16 | Control   | 12 | 12 | 11 |
| 16 | Treatment | 12 | 11 | 9  |
| 17 | Control   | 12 | 12 | 10 |
| 17 | Treatment | 12 | 10 | 8  |
| 18 | Control   | 12 | 15 | 11 |
| 18 | Treatment | 12 | 13 | 8  |
| 19 | Control   | 12 | 12 | 12 |
| 19 | Treatment | 12 | 12 | 10 |
| 20 | Control   | 12 | 12 | 10 |
| 20 | Treatment | 12 | 12 | 10 |

The analysis comparing the predictions of statistical models to those made by human forecasters was pre-registered in [AsPredicted #191059](AsPredicted #191059). In this pre-registration, we provided the predictions made by each of the statistical models and described, prior to the collection of the human forecasts, how forecasters would be recruited and how their forecasts would be analyzed. We note a deviation from the pre-registered predictions for the MLP due to subtle differences in the computing environment used to generate the pre-registered MLP predictions and the fully reproducible code and environment included in this study's repository, which affected the sensitive MLP hyperparameter optimization process. The effect on the model's performance is minimal (the pre-registered predictions have a 0.09 higher RMSE), and we describe the hyperparameters selected by both processes in Section S9. We also note that the pre-registration specified that 80 researchers would be invited to participate, but only 71 were reachable; the remaining researchers either had defunct email addresses or were deceased.

The pre-registration for this study was split into two documents, as we wanted to pre-register the specific predictions made by our models, but the intended prediction task required knowledge of the cooperative outcomes in the control condition. Consequently, we pre-registered the validation configurations to collect the PGG data, then pre-registered the predictions of the models which use the PGG configuration and the outcome in the control condition to predict the outcome in the treatment condition before collecting human forecast data. In retrospect, a stricter pre-registration strategy would have been to combine the two pre-registrations into one, where instead of pre-registering each model's exact predictions we would pre-register the exact Bayesian optimization method (i.e., search space, algorithm/implementation) by which each model's hyperparameters would be selected after collecting the data.

# S2. Data cleansing and processing

## S2.1. Data structure

Data from this study is recorded and aggregated at three levels:

- **Player level:** The decisions (contribution, punishment, reward) made by each player in each round of a PGG. This level of aggregation is used to estimate the average effect of punishment on contributions made by players.

- **Game level:** A "game" is a single group playing a PGG of a given configuration, where a configuration is a combination of PGG design parameters (excluding the availability of punishment). Measures of group efficiency are calculated at the game level.

- **Experiment level:** An "experiment" is a pair of games sharing a PGG configuration, where one game has punishment enabled (treatment) and the other has treatment disabled (control). Games in the validation wave are aggregated at the experiment level by averaging group efficiency in the treatment and control arms, and predictions made by models and humans are made at the experiment level.

## S2.2. Data cleansing & filtration

To prepare the data for analysis, we applied three filters:

- **Technical issue filter:** Three games from both the learning and validation waves were excluded for various technical issues that occurred during data collection.

- **Content-based filter:** In the learning wave, 6 games were excluded for having a multiplier of 1, for which normalized efficiency is undefined. Also excluded from the learning wave was an additional game in which 15 players used the chat function to coordinate a reward ring; by spending all their coins in each round to reward each other, the group achieved gains far exceeding those of a group that fully cooperated in every round (but did not punish/reward). Consequently, the group's relative efficiency was 27, an order of magnitude higher than all other groups in the study, warranting exclusion as an outlier.

- **Insufficient recruitment filter:** A game may start with fewer players than intended for several reasons, such as availability of Prolific workers or long pre-game lobby waits. To ensure that intended (i.e., selected PGG parameters) and actual (i.e., PGGs as executed) configurations are similar, we filter out games missing more than 18% of the intended number of players at the game's start. A player is considered present for the game's start if they complete the first round. This threshold was selected primarily for its simplicity and its impact on games with smaller group sizes; that is, games with 5 or fewer players must always begin with a full group of players. Applying this filter excludes 31 out of 366 games in the learning wave, and 53 out of 470 games in the validation wave.

## S2.3. Measuring group-level cooperation

In the PGG literature, a commonly measured outcome is the "efficiency" of a group, defined as the ratio of $E_{group}$, the group's total earnings (net of all costs and penalties/rewards), to $E_{full\ cooperation}$, the earnings of a hypothetical, fully cooperative group in which every member contributes their full endowment in every round and never punishes or rewards. Mathematically, this is expressed as

$$Efficiency\ =\ \frac{E_{group}}{E_{full\ cooperation}}$$

In our study, this conventional efficiency measure is only used in our human forecasting task, as it is familiar to expert forecasters and appropriate for comparing treatment and control conditions with the same PGG configuration.

However, to compare cooperative behavior across different PGG configurations, we introduce a measure of "normalized efficiency," which addresses differences in the range of possible efficiencies across conditions. For instance, in a PGG where the fund multiplier is 1x and both punishment and reward are disabled, a group's efficiency would always be 1, even if all members defect. By scaling the group's earnings relative to the minimum and maximum possible earnings in each condition, we facilitate direct comparison between cooperative outcomes in PGGs of varying designs. Mathematically, normalized efficiency is expressed as:

$$Normalized\ Efficiency\ =\ \frac{E_{group} - E_{full\ defection}}{E_{full\ cooperation} - E_{full\ defection}}$$

where $E_{full\ defection}$ is the total earnings of a hypothetical group in which all members fully defect (i.e., contribute nothing to the public fund) in every round and never punish or reward. We note that normalized efficiency can exceed 1 when rewards are enabled (e.g., in a setting with full cooperation when participants also reward one another) and can be negative when punishment is enabled (e.g., zero cooperation with punishments). Nonetheless, normalized efficiency acts as a useful, unitless measure that aids comparisons across different game configurations.

To account for potential participant dropout in games, all earnings (both real and hypothetical) are calculated at the round level and then aggregated at the game level. For example, if a PGG began with N players, 1 of whom dropped out in round R, the hypothetical earnings of a fully cooperative/defective group playing this game would assume N players up to round R-1, and N-1 players in rounds including and following round R.

# S3. Robustness checks

## S3.1. Interpretation of the participant dropout exclusion criterion

In our pre-registration, we describe the exclusion criterion as "if the total number of players who drop out is greater than 18% of the intended total number of players, the entire game will be discarded." After we collected the data and implemented this criterion, we realized this was an ambiguous description of our intent. As implemented in the study, we discard a game if the total number of players who drop out *of the game lobby (i.e., pre-game)* is greater than 18% of the intended group size.

An alternative interpretation of the criterion is to discard games where the total number of players who drop out of the game *at any time* is greater than 18% of the intended group size. In Table S2, we reproduce the study's key analyses using this alternative definition, and show that all arising differences are negligible.

**Table S2. Sensitivity of analyses to implemented and alternative exclusion criterion**

| Quantity | Implemented Criterion | Alternative Criterion |
| --- | --- | --- |
| Number of games and paired experiments | Learning: 335 (150 paired)<br>Validation: 417 (40 paired) | Learning: 318 (136 paired)<br>Validation: 382 (40 paired) |
| Contributions (% of endowment) | Learning:<br>0.80 (treatment), 0.73 (control)<br><br>Validation:<br>0.82 (treatment), 0.73 (control) | Learning:<br>0.80 (treatment), 0.73 (control)<br><br>Validation:<br>0.82 (treatment), 0.73 (control) |
| Normalized efficiency | Learning:<br>0.63 (treatment), 0.71 (control)<br><br>Validation:<br>0.68 (treatment), 0.72 (control) | Learning:<br>0.63 (treatment), 0.71 (control)<br><br>Validation:<br>0.67 (treatment), 0.72 (control) |

| Heterogeneity measures | Learning:<br>Q(19) = 29.01, p = 0.066<br>$I^2$ = 0.34<br><br>Validation:<br>Q(19) = 30.29, p = 0.048<br>$I^2$ = 0.47 | Learning:<br>Q(19) = 29.35, p = 0.061<br>$I^2$ = 0.35<br><br>Validation:<br>Q(19) = 32.99, p = 0.024<br>$I^2$ = 0.42 |
|---|---|---|
| Model performance (RMSE) | E-Net: 4.52<br>OLS: 5.22<br>RF: 5.40<br>XGB: 6.33<br>MLP: 5.8 | E-Net: 4.66<br>OLS: 5.72<br>RF: 5.29<br>XGB: 6.21<br>MLP: 8.48 |

## S3.2. Checking for differential inclusion based on PGG design

To check whether a PGG design factors affect whether a game meets our implemented inclusion criterion, we conduct a logistic regression at the game level across both waves of experimentation (N = 836 games in total):

$$ln\left(\frac{p_{inclusion}}{1 - p_{inclusion}}\right) = \alpha + X_{PGG}\beta$$

where $p_{inclusion}$ is the probability of a game meeting the inclusion criterion, α is the intercept, and $X_{PGG}$ is the design matrix of PGG parameters described in Table 1 of the manuscript (with the exception of Peer Incentive Cost, Punishment Impact, and Reward Impact, which are undefined when punishment/reward are disabled). As detailed in Table S3 below, we do not find evidence of PGG design parameters affecting game inclusion in downstream analyses. Although the coefficient for Actor Anonymity is statistically significant (p = 0.036), this finding does not survive correction for multiple comparisons (Benjamini-Hochberg adjusted p = 0.21).

**Table S3. Estimating the effect of PGG design parameters on game inclusion**

| PGG Design Parameter | Coefficient (SE) |
|---|---|
| Group Size | 0.023 (0.026)<br>P = 0.37 |
| Game Length | -0.007 (0.016)<br>P = 0.635 |
| Contribution Type (All or Nothing) | 0.21 (0.253)<br>P = 0.406 |
| Contribution Framing (Opt-out) | -0.142 (0.245)<br>P = 0.56 |

| PGG Design Parameter | Coefficient (SE) |
|---|---|
| MPCR | -0.1 (0.792)<br>P = 0.9 |
| Communication | 0.438 (0.241)<br>P = 0.069 |
| Peer Outcome Visibility | -0.036 (0.256)<br>P = 0.89 |
| Actor Anonymity (Revealed) | 0.515 (0.245)<br>P = 0.036 |
| Horizon Knowledge | 0.299 (0.25)<br>P = 0.232 |
| Punishment Enabled | -0.177 (0.232)<br>P = 0.446 |
| Reward Enabled | -0.02 (0.26)<br>P = 0.937 |

**N = 836 games**

# S4. Selecting PGG configurations through Sobol sequence and random sampling procedures

## S4.1. Sobol sequence (learning wave)

The conditions in the learning wave were generated using a 13-dimensional Sobol sequence with scrambling, as implemented in version 1.10.0 of the SciPy library in Python (Virtanen et al., 2020). By default, SciPy's implementation yields values for each dimension ranging from 0 to 1. The values of the Sobol sequence along each dimension were then scaled (for numerical parameters) or rounded (for binary parameters) to map the Sobol sequence to actual PGG configurations using the ranges provided in Table 1 of the manuscript. As Sobol sequences should be generated in lengths that are powers of 2 (to preserve the sequence's balance properties), a sequence of 256 combinations of PGG parameters (excluding the availability of punishment) was generated, where each of these combinations defines a single experiment and would be executed once with punishment enabled and once without (making punishment the 14th dimension).

It is worth noting that the marginal per capita return (MPCR) cannot be set through the Sobol sequence independently of the group size parameter as this may lead to invalid configurations. For example, an MPCR of 0.2 in a configuration with group size of 4 would imply a multiplier of 0.8, which is a degenerate PGG in which rational participants would not be expected to contribute to the public good. To avoid this, MPCR is determined by an intermediate parameter describing where the MPCR lies between the minimum implied by a multiplier of 1 for that condition's group size (e.g., 0.25 for a group of size 4), and the maximum MPCR of 0.7.

The advantage of using a Sobol sequence in this setting is that the uniformity properties (coverage of the design space) of the sequence are preserved for subsets of points in the sequence. In practice, this means that data collected following the sequence is ideally separated in the design space, even if data collection is terminated early. For the learning wave of experiments, we collected data from the first 181 configurations defined by the Sobol sequence with the exception of 3 configurations that were skipped due to administrative error; the termination of the learning wave was determined by budgetary constraints.

## S4.2. Random sampling (validation wave)

The conditions in the validation wave were generated randomly by independently sampling each parameter from a uniform distribution over the same ranges used in the learning wave's Sobol sequence. The configurations generated for the validation wave were checked against those from the learning wave to ensure that there was no overlap and that the configurations in the validation wave were in fact out-of-sample.

# S5. Participant recruitment, compensation, and onboarding

## S5.1. Participant consent and institutional review

The study was determined to be exempt under Category 3 (Benign Behavioral Intervention) by the Committee on the Use of Humans as Experimental Subjects at the Massachusetts Institute of Technology (Protocol #E-5462). All participants provided explicit consent.

## S5.2. Participant recruitment

A total of 7,100 participants were recruited from a curated panel on Amazon Mechanical Turk (N=604) and from Prolific (N=6,496) and randomly assigned to groups to play synchronous PGGs of varying design.

Initial experimentation in the learning wave began with participants on Amazon Mechanical Turk who were enrolled in a panel of attentive participants curated by the authors. Due to issues with participant availability and the quality of the general (non-panel) population on Amazon Mechanical Turk, the remaining data in the learning wave and the entirety of the validation wave was collected on Prolific. To avoid wasting valuable data, learning wave data from 7 early experiments conducted with the Amazon Mechanical Turk participants was retained.

Participants were recruited from Prolific through the platform's standard sample, filtered to those from the USA, UK, and Canada with self-declared fluency in English, more than 50 previous submissions to Prolific, and a submission acceptance rate greater than 90%. Participants were only allowed to participate once. To facilitate simultaneous recruitment of hundreds of participants within minutes, the tasks on Prolific were configured to disable the default "rate limiting" that Prolific imposes on the distribution of tasks to participants.

## S5.3. Randomization into PGG configurations

Due to server capacity constraints, games were run in batches at a time, such that each batch contained both the treatment and control arm for a given PGG configuration and the total number of players across all configurations in the batch (i.e., number of players simultaneously connected to the server) was approximately 70 players. For each batch, a wave of participants was sent to the server, and each participant was randomly assigned to one of the available configurations.

## S5.4. Participant compensation

Participant compensation was structured as follows:

- **Base pay:** All participants were paid $1.00 for showing up to participate in the study, regardless of their performance. Participants who accepted the task on Prolific but did not proceed through the instructions before their assigned lobby was filled were also paid this amount for showing up on time.

- **Per round pay:** To account for variable game lengths (games could have 1–30 rounds), participants were paid $0.10 for each round they participated in.

- **Performance pay:** Participants were compensated at a rate of $1 per 300 in-game coins they earned.

Participants were also granted "lobby pay" at a rate of $15/hr for any time spent in the pre-game lobby waiting for other players to arrive. However, unlike base, per round, and performance pay, participants were not informed that they would receive this lobby pay ahead of time to keep the perceived scale of the incentives as consistent as possible across participants. The minimum total earnings rate was $8/hr, and the median was $17/hr.

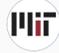

**Figure S1. Prolific recruitment text shown to PGG participants**

## S5.5. Participant onboarding

Upon joining the experiment, participants were first guided through an annotated version of the actual in-game interface to ensure that they understood the game and were familiar with its interface. Each stage of the game was explained separately, with a comprehension check required to proceed. This walkthrough was tailored to the configuration that the participants were randomly assigned to; for example, if a configuration did not allow communication between participants, the chat window would not be rendered or explained, and vice versa. At the end of the interface walkthrough, participants were asked to (1) confirm that they understood that the game may run for any duration from 5–50 minutes and that quitting or idling meant they forfeited their bonus payments, and (2) submit the completion code on Prolific *before* proceeding to the game lobby in order to relieve any participant anxieties about the task "timing out" on the Prolific platform. Example screenshots of the walkthrough are included below (for a walkthrough of a full-featured configuration including chat, punishment, and reward mechanisms, please see the study repository referenced in Section S1):

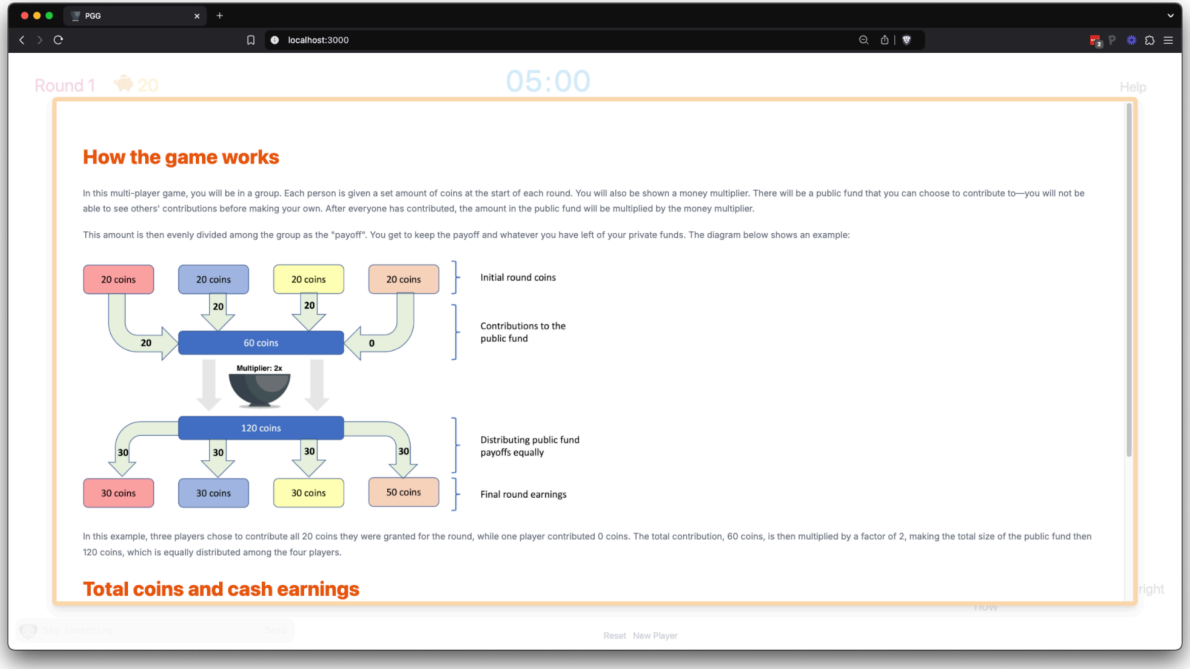

**Figure S2. Interface walkthrough example (game rules)**

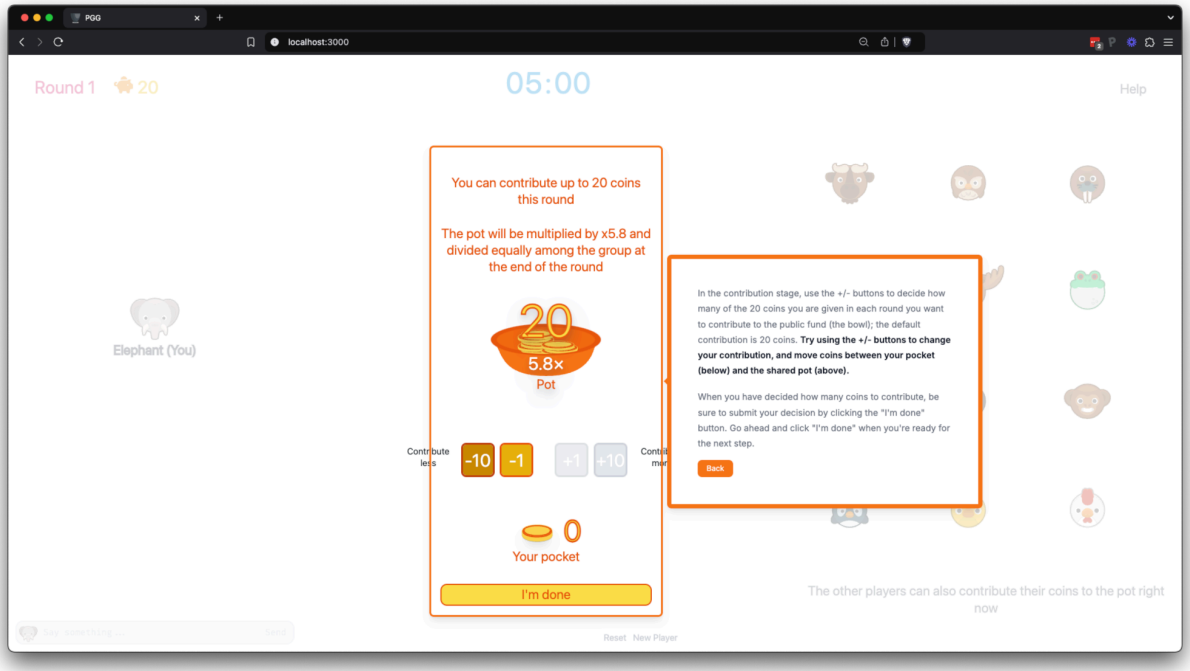

**Figure S3. Interface walkthrough example (contribution interface)**

# S6. Interface design and implementation details

The experimental interface was implemented using Empirica (https://empirica.ly/) (Almaatouq et al., 2021), an open-source software designed for integrative experimentation through multi-participant online games with real-time interactivity. Here, we illustrate how each PGG design parameter is reflected in the interface, and discuss implementation details related to randomization, idle detection, and in-game calculations.

## S6.1. PGG game structure

Our PGG is implemented as a series of rounds, where each round is composed of three consecutive stages:

1. In the **contribution** stage (Figure S4), players are granted a per round endowment and decide how much of the endowment to contribute to the public fund.

2. In the **redistribution** stage (Figure S5), players are shown the coins contributed by each player, as well as the sum of all contributions, and the amount redistributed to each player. During this stage, players may punish and/or reward other players, depending on the configuration of the PGG.

3. In the **round summary** stage (Figure S6), players view a summary of their earnings, including their retained endowment, share of the public fund payoff, and punishments/rewards that were given/received; depending on the PGG configuration, they may also view this information for other players. No actions are taken during this stage. After the *round summary* stage is complete, another round begins starting from the *contribution* stage.

At the end of each round, a player's net earnings/losses for that round are added to the player's cumulative earnings for the game (a.k.a. their "piggy bank", shown in the top left corner of the interface). These cumulative earnings are the money that players use to pay for punishments and rewards they choose to impose in the redistribution stage. To allow players to apply punishments and rewards in the first round, each player starts the game with 20 coins in their piggy bank; this is a one-time grant at the beginning of the game, unlike the per round endowment.

## S6.2. Reflecting PGG design parameters in the interface

Below, we explain each of the 14 PGG design parameters and illustrate how they are reflected in the game interface:

**Group Size:**
- Definition: Number of players in the game.
- Interface: Other players are always shown in a grid on the right side of the screen, as highlighted in Figure S4A.
- Notes: When a player drops out mid-game, they no longer appear in the user interface for other players, and consequently cannot receive any rewards or punishments. The public fund is divided among the remaining players, without adjusting the multiplier to maintain the intended MPCR.

**Game Length and Horizon Knowledge:**
- Definitions: Game Length is the number of rounds in the game, while Horizon Knowledge describes whether players know the total length of the game.
- Interface: The current round ("Round X") is always displayed. When players are given knowledge of the horizon, the ("of Y") is added, as shown in Figure S4B.

**Contribution Type:**
- Definition: Whether players can contribute any amount or must choose between contributing all or none of their endowment.
- Interface: When the contribution type is "variable", players can add or subtract any number of coins to their contribution (up to the per round endowment) using buttons to add or subtract 1 or 10 coins, as illustrated in Figure S4C. When the contribution type is set to "all or nothing", the only contribution buttons available are +/- 20 coins.

**Contribution Framing:**
- Definition: Whether each player's endowment starts in their private account (opt-in) or in the public fund (opt-out).
- Interface: When the contribution is framed as "opt-in", players will see 20 coins in their pocket (Figure S4D1) and 0 coins contributed by them to the public fund (Figure S4D2). When the contribution is framed as "opt-out", the amounts are reversed.

**Marginal per capita return (MPCR):**
- Definition: Marginal per capita return on contributions, defined as the fund multiplier divided by group size.
- Interface: The MPCR is reflected in the multiplier shown to players (Figure S4E).

**Communication:**
- Definition: Whether players can communicate through a chat window during the game.
- Interface: When communication is enabled, an expandable chat window is added (Figure S4F).

- Notes: The scope of the chat window spans the entire game; that is, messages do not disappear between rounds.

**Punishment/reward availability, cost, and impact**
- Definitions: Punishment and reward are manipulated independently of each other; that is, a configuration may have one, both, or none of these peer incentive mechanisms enabled. Each mechanism has a cost to the player using the mechanism (coins paid per unit of punishment or reward) and an impact on the targeted player (a multiple of the total number of coins used to punish or reward them).
- Interface: During the redistribution stage, when punishment and/or reward is enabled, players will see "-" buttons above peer avatars to deduct coins (one click per punishment imposed) and "+" buttons to grant coins (one click per reward); see Figure S5A. When punishment is enabled but reward is disabled, "+" allows players to reduce the amount of punishment they would like to deliver, but it will not increase the coins above 0 (since reward is disabled). The "-" button behaves similarly when reward is enabled and punishment is disabled. The costs/impacts of the punishment/reward mechanisms are communicated as shown in Figure S5B, as applicable to the PGG's configuration.
- Notes:
  - The cost per unit of punishment and the cost per unit of reward are held equal in all PGG configurations, and reflected in a single parameter called "Peer Incentive Cost".
  - The total cost of punishments and rewards imposed by a player in a given round is deducted from the player's cumulative earnings in the game thus far.

**Peer Outcome Visibility**
- Definition: Whether players can see summaries of others' earnings and punishments/rewards received in each round during the round summary stage.
- Interface: In the round summary stage, players are shown summaries of their gains and losses from various sources through information surrounding their avatar (Figure S6A) and detailed cards that appear when hovering over any player's avatar, including their own (Figure S6B). When peer outcomes are hidden, peer avatars are rendered without surrounding information, and information cards do not appear when hovering over peer avatars.
- Notes: This parameter only applies to the round summary stage; contributions made by peers are always publicly visible in the redistribution stage.

**Actor Anonymity**
- Definition: Whether the identity of players who punish or reward others is revealed.
- Interface: When identities are revealed, players can see the identity of players who have punished or rewarded them and others in the detailed information card that appears when hovering over any player's avatar (Figure S6C). When identities are hidden, players can see the total amounts of punishment and reward in the information cards, but they are not itemized by the acting player.

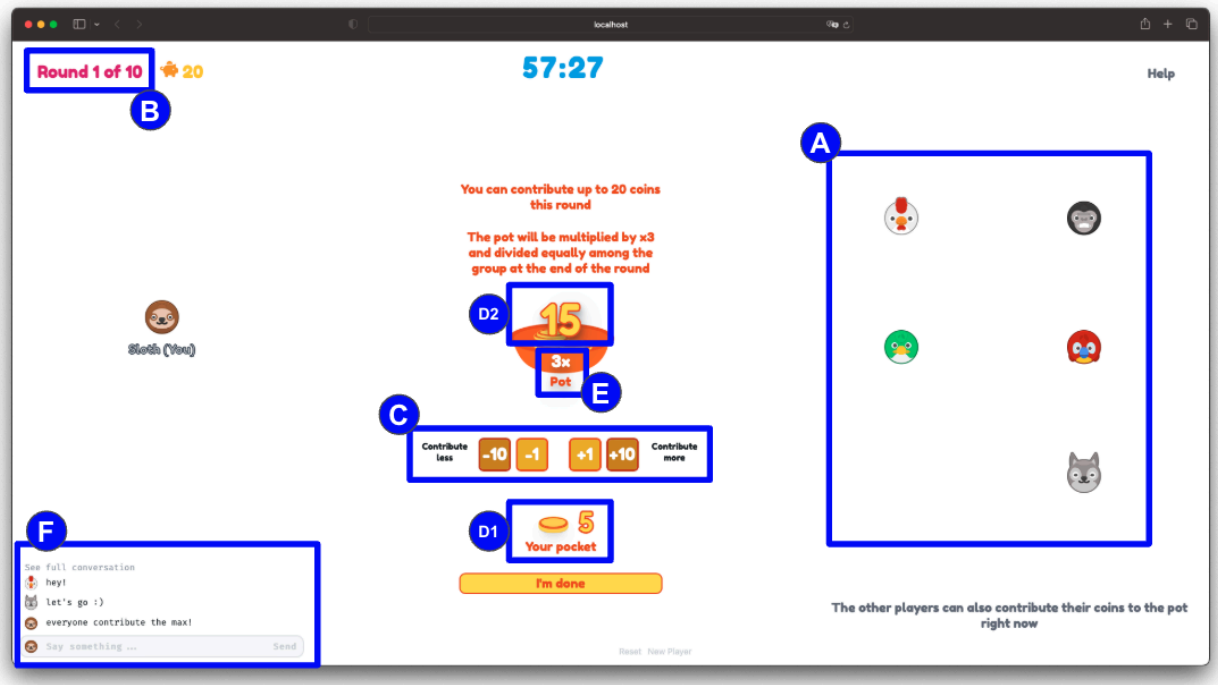

**Figure S4. "Contribution"-stage interface.** In the contribution stage, players decide how much of their per round endowment to contribute to the public good. Factors manipulated in this stage are the number of players (A), the number of rounds and visibility of the game horizon (B), whether contributions can vary in amount or can only be "all or nothing" (C), whether contribution is opt-in or opt-out (D1,D2), the multiplier applied to public fund contributions (E), and the availability of a chat window (F).

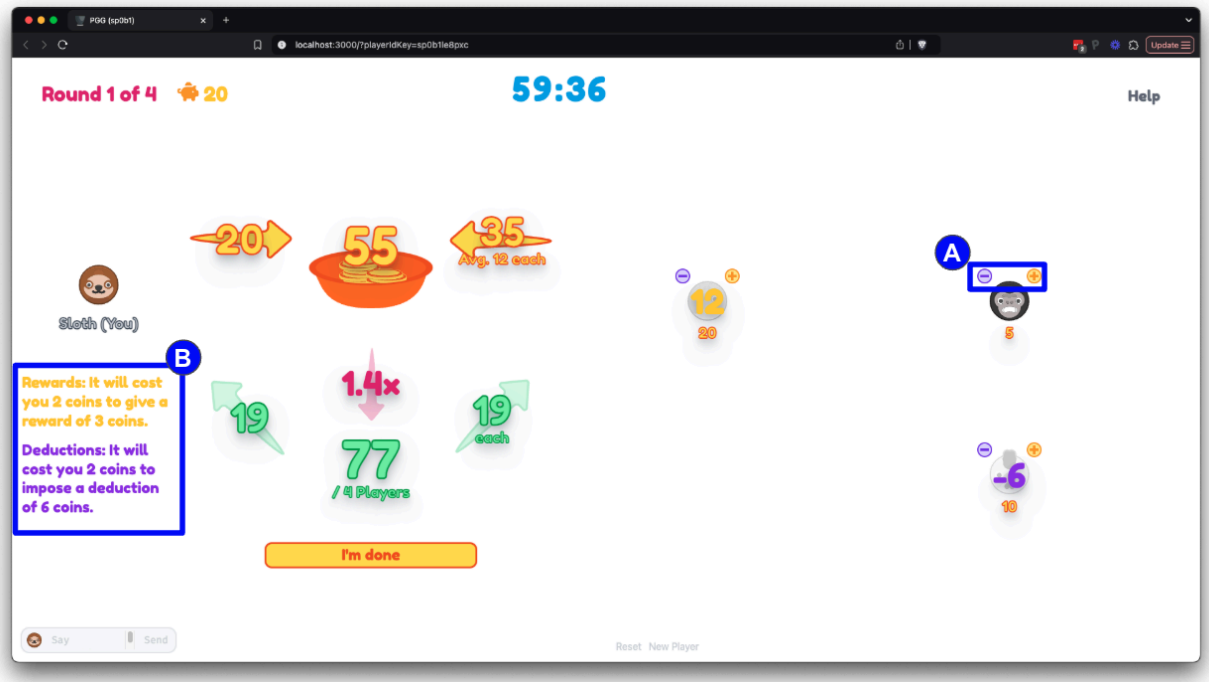

**Figure S5. "Redistribution"-stage interface.** In the redistribution stage, players observe the contributions made by all players and the division of the public good. The key factors manipulated in this stage are the availability of punishment and reward mechanisms (A) and the associated costs/impacts (B), as applicable to the PGG's configuration.

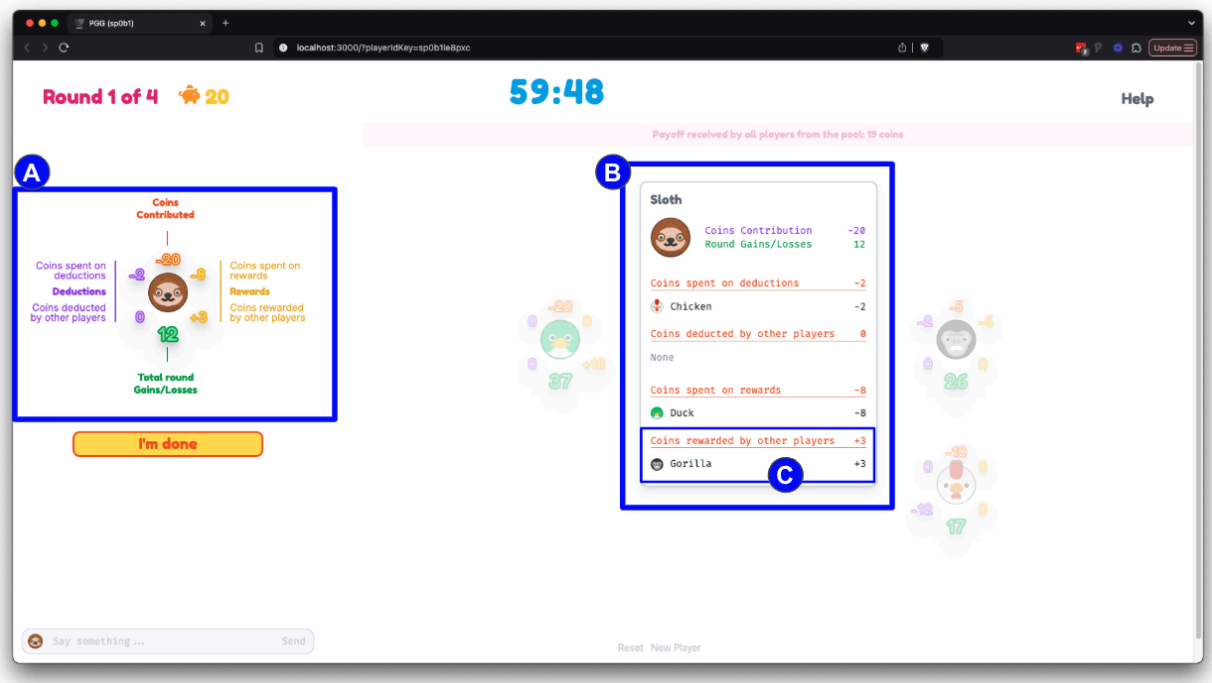

**Figure S6. "Round summary"-stage interface.** In the round summary stage, players are shown high-level (A) and detailed (B) summaries of their performance in the round, as applicable to the PGG's configuration.

In addition to the 14 actively manipulated PGG parameters described above, the following parameters were held constant across all games in both the learning and validation waves:

- **Stage duration:** Each of the game's three stages had a maximum duration of 45 seconds; that is, each round of the game took at most 135 seconds. However, if all players clicked the button confirming that they were ready to proceed from a given stage, the game would immediately move to the next stage.

- **Per round endowment:** In all games, players were granted 20 coins in each round and had to decide how much of this per round endowment (as opposed to their cumulatively earned coins) to contribute to the public good for that round.

## S6.3. Idle player detection and removal

To ensure active participation throughout the duration of the games, we implement two measures to detect and remove inactive players.

The first measure is a *ready check*: after all players have completed the interface walkthrough, they must click a button to confirm that they are ready to begin; players who fail to press the button in time are removed before the start of the game. After all players have completed the interface walkthrough, they enter a lobby where they wait for other players assigned to the same configuration to complete the walkthrough. After the required number of players has joined the lobby, each player must click a button to confirm that they are ready to begin; players who fail to press the button within 20 seconds are removed before the start of the actual game. This mechanism prevents the first stage of the game from being affected by players who may have decided not to participate or stepped away during the lobby period, and minimizes player confusion caused by removing such players mid-game. This mechanism was only implemented approximately halfway into the learning wave of experimentation, but was in place for all games in the validation wave.

The second measure checks for interaction with the interface during the game. Players who do not interact with the interface for more than 50 seconds are shown a pop-up asking them to confirm that they are present. If they fail to do so within an additional 30 seconds, they are removed from the game. Additionally, if a player is detected to be offline for more than 3 stages of the game, they are removed.

## S6.4. Adapting to participant dropout

Considering the size of the groups and duration of the games in this study, participant dropout (e.g., leaving mid-game or abandoning the lobby after assignment) is to be expected. Accordingly, when a participant drops out, the interface adapts by removing their avatar from the game (i.e., no information is shown about them, and they cannot be targeted for punishment or reward) and only considering the number of active players when dividing the public good among players. However, we note that the multiplier applied to the public fund does not change mid-game in order to maintain the intended MPCR; in this sense, factors such as the MPCR should be considered as an "intent to treat" variable with the value assigned to the configuration.

## S6.5. Rounding up/down when dividing the public good

In cases where the value of the public good (total contributions multiplied by the contribution multiplier) is not divisible by the number of players, the per player share of the public good is rounded to the nearest integer.

# S7. Estimating the average effect of punishment on contributions and normalized efficiency

To account for the hierarchical structure of contribution-level data, the average effect of punishment on contributions is estimated in a mixed effects linear model with random intercepts at the player level (players make multiple contribution decisions, one per round) and the game level (players are grouped by games):

$$Contribution_{irg} = \alpha + \beta_{punishment} Punishment_g + \nu_i + \gamma_g + \varepsilon_{irg}$$

where $Contribution_{irg}$ is the proportion of the per round endowment contributed by player $i$ in round $r$ of game $g$, $\alpha$ is the fixed intercept, $\beta_{punishment}$ is the fixed effect of punishment on contributions, $\nu_i$ and $\gamma_g$ are random intercepts for player $i$ and game $g$ respectively, and $\varepsilon_{irg}$ is the observation-level error term. If a player is removed before the end of a round, their contribution for that round is not included in the regression.

As a robustness check, we estimate the effect of treatment on contributions through a two-sided tobit regression censored at 0 (because negative contributions are not possible) and 1 (because contributions are capped by the per round endowment). For this specification, the dependent variable is the average contribution within a game across all rounds and players, eliminating the need for player and game random intercepts. As summarized in Table S4, the estimates from both specifications are similar.

**Table S4. Estimating the average effect of punishment on contributions**

|  | Wave / Model | | | |
|---|---|---|---|---|
|  | Learning | | Validation | |
|  | Mixed effects | Game-level tobit | Mixed effects | Game-level tobit |
| **Intercept** | 0.71 (0.01) | 0.71 (0.01) | 0.71 (0.01) | 0.71 (0.01) |
| **Punishment** | 0.066 (0.02) | 0.063 (0.02) | 0.077 (0.02) | 0.074 (0.02) |
| **Number of players** | 3,618 | | 3,482 | |
| **Number of games** | 335 | | 417 | |
| **sd(players)** | 0.23 | | 0.22 | |
| **sd(games)** | 0.15 | | 0.14 | |
| **N** | 53,973 | 335 | 48,370 | 417 |

On the other hand, normalized efficiency is a group-level measurement and groups are independent of one another, so the average effect of punishment on normalized efficiency is estimated directly via OLS:

$$NormalizedEfficiency_g = \alpha + \beta_{punishment} Punishment_g + \varepsilon_g$$

**Table S5. Estimating the average effect of punishment on normalized efficiency**

|  | Wave | |
|---|---|---|
|  | Learning | Validation |
| **Intercept** | 0.714 (0.022) | 0.720 (0.016) |
| **Punishment** | -0.089 (0.032) | -0.043 (0.022) |
| **N** | 335 | 417 |

# S8. Detecting and measuring heterogeneity in the effect of punishment

## S8.1. Segmenting the design space in the learning wave to detect heterogeneity in the effect of punishment on normalized efficiency

To analyze the heterogeneity underlying the effect of punishment on normalized efficiency, we rely on Cochran's Q and $I^2$, two measures that depend on the variance in estimated treatment effects across experimental conditions. While the experiments in the validation wave had 8–12 trials each to facilitate this analysis, the experiments in the learning wave were designed for breadth, not precision, and only had one trial each.

Therefore, to facilitate the visualization in Figure 2 from the manuscript, we cluster configurations in the learning wave based on their design parameters. Using the k-means implementation in version 1.5.1 of the scikit-learn package in Python (Pedregosa et al., 2011), we split the learning set experiments into 20 clusters—this number was chosen to mirror the number of experiments in the validation set. Before clustering, PGG design features with larger scales (the number of players and rounds as well as the peer incentive cost and punishment impact) were min-max scaled to lie between 0 (corresponding to the minimum value) and 1 (corresponding to the maximum value), bringing them into the same scale as the MPCR and boolean features, and the reward impact parameter was dropped as it is not applicable when reward is disabled.

Table S6 presents the clusters generated for Figure 2 in the manuscript, including the number of configurations and games included in each, and the estimated effect of punishment on normalized efficiency (difference in means) within each cluster. To reiterate, the main purpose of clustering the learning wave games is visualization; the heterogeneity measurements in the validation wave, where each experiment is pre-registered and has 8–12 trials, are the more appropriate and reliable representation of heterogeneity in the effect of punishment. Intuitively, the heterogeneity measured within the clusters of learning wave experiments is indeed sensitive to the coarseness of the clustering, among other hyperparameters; with too few clusters, the estimate in each cluster effectively approximates the average treatment effect across the design space, and with too many clusters (and sufficient precision) even miniscule variation can be detected.

**Table S6. Data points and punishment effect estimates within each learning wave cluster**

| Cluster ID | Unique configurations | Games in treatment | Games in control | Within-cluster punishment effect (SE) |
|---:|---:|---:|---:|---|
| 0 | 11 | 11 | 11 | 0.101 (0.087) |
| 1 | 4 | 3 | 4 | -0.01 (0.057) |
| 2 | 8 | 8 | 9 | 0.246 (0.072) |
| 3 | 12 | 10 | 11 | -0.031 (0.08) |
| 4 | 11 | 10 | 11 | 0.032 (0.082) |
| 5 | 3 | 4 | 4 | 0.029 (0.096) |
| 6 | 11 | 9 | 11 | -0.051 (0.062) |
| 7 | 5 | 5 | 5 | -0.019 (0.136) |
| 8 | 9 | 9 | 9 | -0.107 (0.1) |
| 9 | 11 | 11 | 10 | -0.015 (0.095) |
| 10 | 6 | 6 | 7 | -0.341 (0.224) |
| 11 | 14 | 16 | 16 | -0.244 (0.106) |
| 12 | 11 | 12 | 10 | -0.24 (0.177) |
| 13 | 10 | 11 | 11 | -0.056 (0.085) |
| 14 | 6 | 5 | 6 | -0.117 (0.134) |
| 15 | 6 | 5 | 6 | -0.25 (0.193) |
| 16 | 10 | 10 | 8 | -0.307 (0.308) |
| 17 | 10 | 9 | 10 | -0.094 (0.136) |
| 18 | 5 | 4 | 5 | 0.009 (0.13) |
| 19 | 7 | 6 | 7 | -0.083 (0.144) |

# S9. Predictive model selection, performance baselines, and uncertainty quantification

## S9.1. The prediction task

In this study, the prediction task is to predict the average group efficiency under treatment (with punishment) given the design parameters of a PGG experiment and the average group efficiency under control (without punishment enabled). Accordingly, game-level data is converted to experiment-level data by averaging group efficiency within the experiment's treatment and control arms. Because the availability of punishment is the focal treatment, the peer incentive cost and punishment impact are relevant to all experiments. However, not all experiments include the reward mechanism, so reward impact is not included as an input feature. All 13 other PGG design parameters are used as inputs to the models.

Among the 335 valid games in the learning set, there are 150 matched pairs for which the treatment and control arms were both valid and can consequently be used for the prediction task—we label this matched dataset $D_{learning}$. For the validation set, since each configuration was executed several times, there are no unmatched configurations and $D_{validation}$ has 20 configurations.

Models were primarily evaluated by measuring the root mean squared error (RMSE) of their predictions of average group efficiency under punishment in the validation wave, the data for which was held out during model training. A secondary measure of performance is out-of-sample $R^2$ (referred to simply as $R^2$), defined as

$$R^2 = 1 - \frac{\sum_i (y_i - \hat{y}_i)^2}{\sum_i (y_i - \bar{y}_{learn})^2}$$

where $i$ denotes the index of the experiment in the validation wave and $\bar{y}_{learn}$ is the average group efficiency under punishment in the learning experiments.

## S9.2. Bayesian optimization to select model parameters

Five models were trained using $D_{learning}$ to predict the outcomes in $D_{validation}$: ordinary least squares regression (OLS), elastic net regularized regression (E-Net) (Zou & Hastie, 2005) with two-way interactions, a random forest (RF) (Breiman, 2001), XGBoost (XGB) (Chen & Guestrin, 2016), and a multilayer perceptron (MLP, also known as a dense neural network).

The hyperparameters of the E-Net, RF, MLP, and XGB models were selected by Bayesian optimization (implemented in Python with the gp_minimize function from version 0.10.2 of the scikit-optimize library) (Head et al., 2021), where the objective function was minimizing the 10-fold cross-validated mean squared error on $D_{learning}$, calculated by aggregating predictions across all folds. For each model, the optimization ran for 50 iterations—all other gp_minimize parameters were left to their default values.

Table S7 lists the specific implementation of each model, the search space of hyperparameters, the best parameter values found, and the cross-validated and out-of-sample RMSEs. The search space for each model's parameters was set by a best-effort attempt to identify the parameters most relevant to overfitting (given the small size of $D_{learning}$) and setting ranges that seemed reasonable given the size of the training data.

The E-Net and MLP model pipelines included a standardization step applied to the features (i.e., not the target variable) of $D_{learn}$ to avoid issues with the L1/L2 regularization of the E-Net across variables of different scales and to improve the stability of the MLP optimization. The input to the E-Net was further preprocessed to include 2-way feature interactions. All other models used raw feature values.

The MLP was implemented as a dense neural network with the same number of units in each layer to simplify the hyperparameter search. Between the input layer (with units equal to the dimension of the input) and the output layer (a single unit), each layer uses a rectified linear unit (ReLU) activation and is followed by a dropout layer. The MLP was optimized to minimize mean squared error by adaptive moment estimation (Adam), with the learning rate set by the hyperparameter search. While the number of epochs was set as a search parameter, the batch size was kept constant at 32. As noted in Section S1, the hyperparameters used for the pre-registered predictions were optimized on a different computing environment than the fully reproducible results in Table S7; the pre-registered MLP differs only in having 34 units per layer instead of 21.

**Table S7. Statistical model parameter spaces, optimized parameter sets, and RMSEs**

| Model | Parameter | Search Space Type(min,max) | Best Value | RMSE (CV) | RMSE (OOS) |
|---|---|---|---|---|---|
| LinearRegression ("OLS") scikit-learn 1.5.1 | NA | NA | NA | 13.16 | 5.22 |
| ElasticNet ("E-Net") scikit-learn 1.5.1 | alpha | Real(0.001, 0.1, "log-uniform") | 0.07 | 12.94 | 4.52 |
| | l1_ratio | Real(0,1) | 0.15 | | |
| RandomForestRegressor ("RF") scikit-learn 1.5.1 | max_depth | Integer(2,6) | 4 | 12.56 | 5.4 |
| | min_samples_leaf | Integer(1,20) | 12 | | |
| XGBRegressor ("XGB") xgboost 1.7.6 | n_estimators | Integer(2,15) | 10 | 12.95 | 6.33 |
| | max_depth | Integer(2,6) | 2 | | |
| | gamma | Real(0,1) | 0.0053 | | |
| KerasRegressor ("MLP") scikeras 0.13.0 | n_layers | Integer(2,5) | 2 | 12.53 | 5.8 |
| | n_units_per_layer | Integer(20,80) | 21 | | |
| | learning_rate | Real(1e-6,1e-1, "log-uniform") | 0.012 | | |
| | dropout_rate | Real(0,0.3) | 0.3 | | |
| | epochs | Categorical([25, 50,100]) | 100 | | |

All results reported in this study were computed on an Apple M2 Pro processor with the macOS Sonoma 14.5 operating system. Holding the hardware and Python environment constant, the MLP model selection procedure produces a different model when run on macOS Sequoia 15.3; given that all other model selection procedures are unaffected, this is likely due to changes in system-level computational libraries between the operating system versions. On macOS 15.3, the MLP hyperparameters selected by the procedure are 5 layers, 80 units per layer, and a learning rate of 0.006, while the dropout rate and number of epochs remain the same. With this set of parameters, the cross-validated RMSE is 13.38, and the out-of-sample RMSE is 5.07.

## S9.3. Statistical baselines of model performance

To contextualize the predictive performance of the statistical models, we compare them to three statistical baselines:

1. Predicting that punishment has no effect (i.e., the average efficiency under control and treatment are equal), which achieves an RMSE of 7.13.

2. Using the average treatment effect (ATE) in the learning wave and average group efficiency under control in the validation wave to predict average efficiency under treatment, which achieves an RMSE of 6.79. The fact that all models outperform this baseline can also be considered evidence supporting the existence of treatment effect heterogeneity.

3. Using the average efficiency under treatment in the learning wave as a constant prediction for all configurations in the validation set, which achieves an RMSE of 6.61.

## S9.4. Bootstrapping to estimate the uncertainty of model performance over validation experiments

The predictive performance of our models on the 20 experiments in the validation wave is an unambiguous quantity, simply calculated as the RMSE between a model's predictions and the mean efficiency under treatment in the validation wave experiments. However, the PGG parameters in the validation wave were randomly sampled from a distribution of conditions, and we would also like to quantify the uncertainty in the model performance arising from this sampling of conditions. To do so, we used a nonparametric bootstrap procedure with 1000 resamples. For each resample, we sample 20 learning wave experiments with replacement from the original set of 20 experiments. We use these sampled experiments to calculate the RMSE of each model/baseline.

The results of this bootstrapping procedure can be interpreted in two ways. To construct confidence intervals (CIs) for the RMSE of a single model, we calculate the relevant quantiles in the model's bootstrapped RMSE values. However, to construct CIs for the difference in RMSE between two models or baselines, we should calculate the quantiles on the bootstrapped distribution *of differences* between the two models, rather than comparing the individual CIs of model RMSE. For each model, the 95% CI of its RMSE and the reduction in RMSE relative to the baseline prediction of the mean efficiency under treatment in the learning set is reported in Table S8; out of all models and the wisdom of both lay and expert crowds, the E-Net model is the only predictor that consistently outperforms the baseline (i.e., $RMSE_{model}$ / $RMSE_{baseline}$ < 1).

**Table S8. Bootstrapped model and forecaster performance**

| Model | RMSE 95% CI | $RMSE_{model}$ / $RMSE_{baseline}$ 95% CI |
|---|---|---|
| E-Net | [3.35, 5.69] | **[0.53, 0.90]** |
| OLS | [3.93, 6.52] | [0.57, 1.14] |
| RF | [3.55, 7.39] | [0.55, 1.11] |
| XGB | [4.19, 8.52] | [0.63, 1.34] |
| MLP | [4.16, 7.33] | [0.63, 1.21] |
| WoC Prolific | [4.59, 8.08] | [0.68, 1.35] |
| WoC SSPP | [4.61, 8.77] | [0.68, 1.36] |

# S10. Human forecasting task implementation, recruitment, and evaluation

## S10.1. The prediction task

To benchmark the performance of statistical models, human forecasters were recruited to predict the average efficiency of groups playing PGGs with punishment enabled, given the configuration of the PGG and the average efficiency of groups playing the same PGG with punishment disabled, for the 20 different PGG experiments in the validation wave. For the prediction task, we chose efficiency as the predictive target because it is the most commonly used measure of cooperation in the PGG literature, would be familiar to our academic forecasters, and is appropriate for within-experiment comparisons between control and treatment conditions

Upon accepting the task, forecasters were presented with an introduction to PGGs and explanations of the experimental setup and efficiency measure. They were then given instructions about the prediction task, including a note that they would be evaluated by the squared error of their predictions (Figure S7). Before proceeding to the prediction task (an example of which is shown in Figure S8), participants completed a brief check of their comprehension of PGG mechanics and the group efficiency measure. Participants were told that the information presented before beginning the task could be accessed at any time during the task through a publicly available Google Doc (for the PDF version of this document, please see the study repository referenced in Section S1).

While there was no time limit, participants were required to make at least 10 out of 20 predictions for the task to be considered complete and to receive the completion code upon exit; 1 out of 54 expert forecasters and 12 out of 512 forecasters recruited from Prolific failed to make at least 10 predictions. As pre-registered, 2 predictions from lay people were excluded for being out of the range [-0.2, 1.2]. The median duration spent in the survey (from intro to exit) by experts and laypeople was 20 and 19 minutes, with a median time per prediction of 29 and 28 seconds, respectively.

All forecasters were informed that they would be judged by the squared error of their predictions; while SSPP forecasters were uncompensated, Prolific workers were incentivized with a base pay of $2.00 and a maximum bonus of $0.50 for each prediction based on prediction accuracy (for a total of $10.00 in potential performance bonuses).

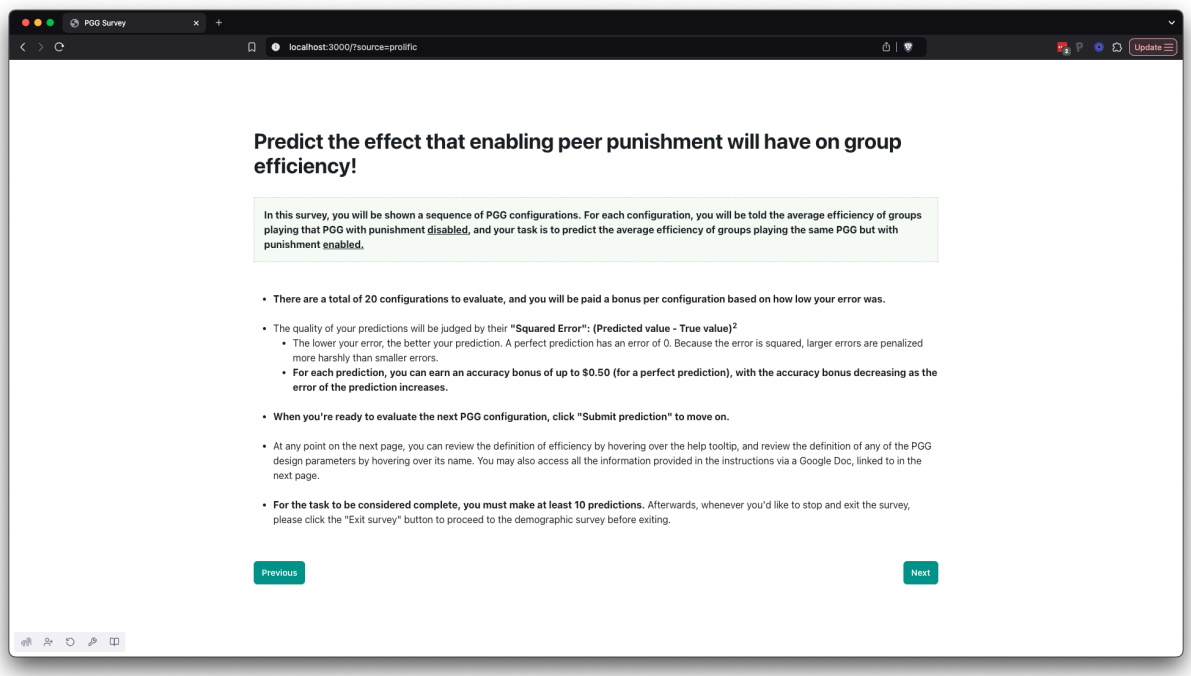

**Figure S7. Prediction task instructions.** All participants were notified that they would be evaluated by the squared error of their predictions. Any information about compensation was only shown to Prolific participants.

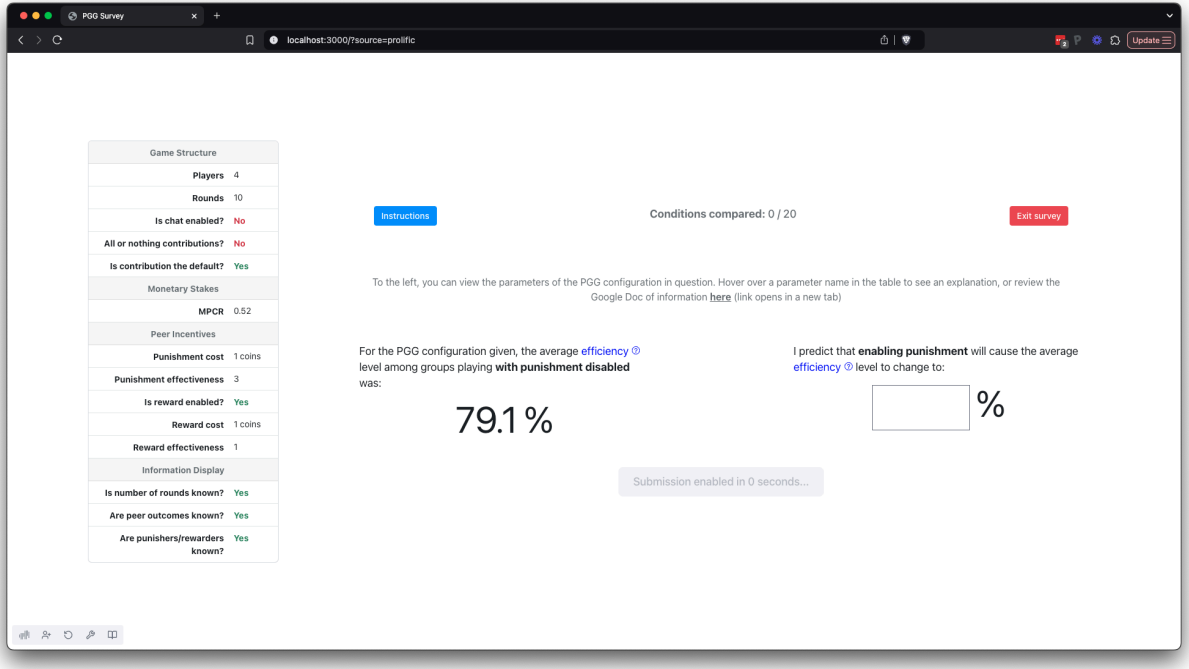

**Figure S8. Prediction task interface.** For each of the 20 validation experiments, participants were shown the experiment's configuration in a table on the left, as well as the average group efficiency in the control (no punishment) arm of the experiment, and asked to submit their prediction of the average group efficiency with punishment enabled. At any point, participants could refer to the full introduction to the task, PGG design parameters, and definition of efficiency. Participants were allowed to complete the task after submitting at least 10 predictions.

## S10.2. Expert recruitment

Through the Social Science Prediction Platform (SSPP) (DellaVigna et al., 2019), a total of 53 academics completed the prediction task. Out of 71 authors of PGG papers who were directly invited to participate, 6 authors completed the prediction task. The remaining 47 academics came from the SSPP's general pool of researchers in related domains (e.g., economics, game theory, psychology). Participants from this pool of academics were not directly incentivized for our prediction task, but some respondents from SSPP's general pool may receive payment from SSPP to participate in surveys posted to the platform.

## S10.3. Layperson recruitment

A total of 500 participants completed the forecasting task from Prolific. Among the participants, 250 of them were Prolific workers who had participated in the learning or validation waves of the experiment, and 250 were naive participants who had never encountered our study. Notably, the survey had a relatively high return rate (deciding not to participate in the task after viewing the details) of approximately 40%; this is likely due to the complexity of the instructions and materials explaining the PGG and prediction task. Consequently, their predictive performance should be interpreted with the recognition that the participants who continued were likely more willing to engage with complex tasks and may not be representative lay people. An additional 12 participants made fewer than 10 predictions; these were thus included in the "wisdom of crowds" estimate but excluded from analysis of individual performance, as pre-registered.

Prolific workers were incentivized with a base pay of $2.00 and a maximum bonus of $0.50 for each prediction based on prediction accuracy. For each configuration, accuracy bonuses for Prolific workers were calculated based on the decile of their prediction's squared error, with the best 10% of forecasters earning the maximum $0.50 that can be earned for a single prediction, and the following deciles earning $0.15, $0.04, and $0.01 per prediction—predictions below the 60th percentile were not bonused. The total accuracy bonus earned by a participant was then calculated by summing across all their predictions. The median bonus pay earned was $1.88.

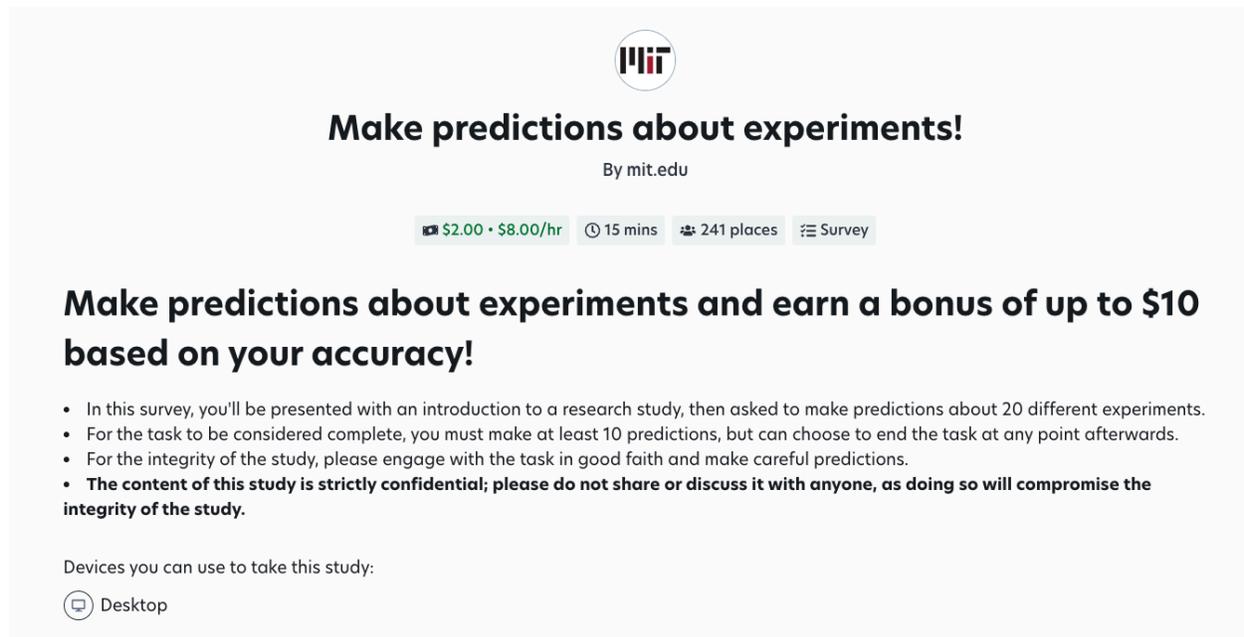

**Figure S9. Prolific recruitment text shown to forecasting task participants**

## S10.4. Individual-level forecaster performance analysis

As pre-registered (AsPredicted #191059), we evaluate the individual performance of models and forecasters by z-scoring the standard error of their predictions within each validation wave configuration. In contrast to direct evaluation by RMSE, standardizing squared error within configurations allows us to both account for the inherent difficulty of predicting the outcome for each configuration and evaluate the predictive power of forecasters who did not complete the full survey (but made at least 10 predictions, as pre-registered) by averaging their z-scored error across configurations.

As illustrated in Figure S10, the E-Net model outperformed all models and forecasters with a mean z-scored squared error of -0.336; that is, the E-Net model's squared error on a given task is, on average, 0.336 standard deviations lower than the squared error of the models and forecasters in our study. The next best source of predictions was a layperson who achieved a mean z-scored squared error of -0.334, and the worst of the statistical models (XGB) was better than 97% of all forecasters.

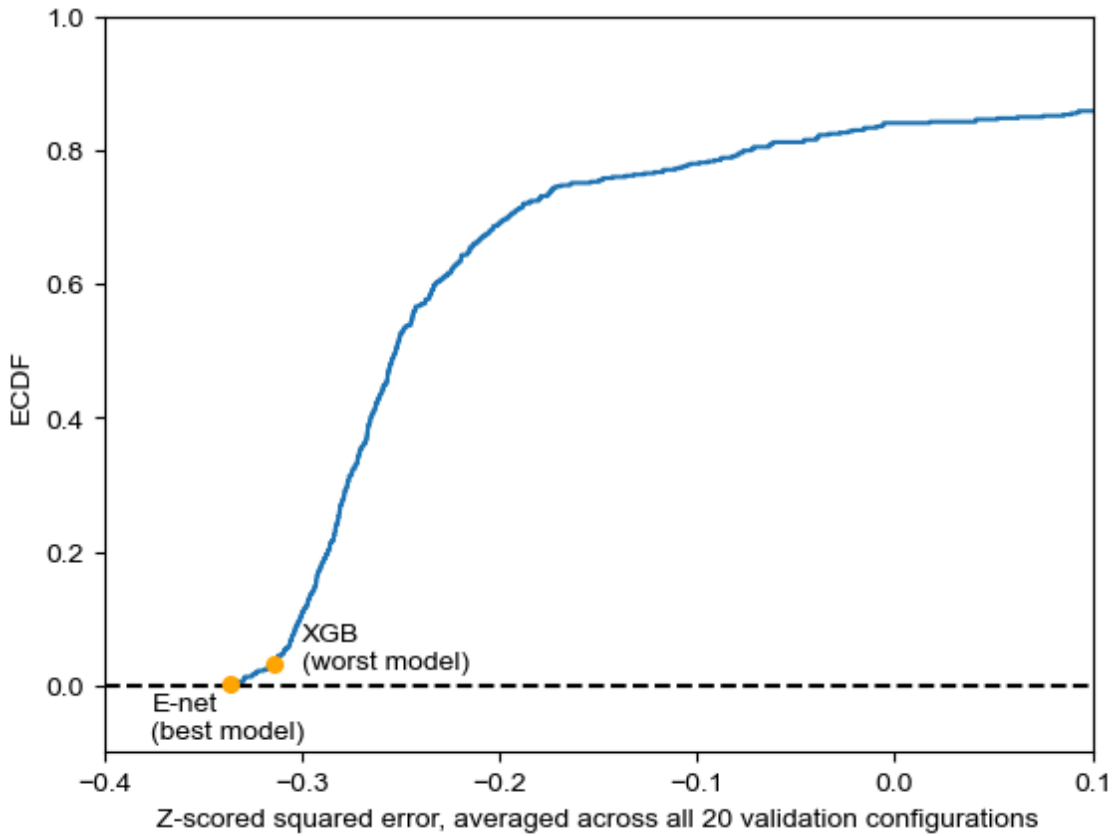

**Figure S10. Distribution of model and individual-level human forecasting performance.** To measure individual-level performance, prediction error is first z-scored within each validation wave experiment. A model or human's individual-level performance on the entire task is then summarized as the mean z-scored error across all 20 validation wave experiments (i.e., a lower value indicates better predictive performance). The best statistical model (E-Net) outperformed all other models and all expert and lay human forecasters, while the worst statistical model (XGB) outperformed 97% of forecasters. The x-axis is truncated for clarity.

# S11. Feature importance

## S11.1. Feature importance measures and implementation

To gauge the importance of each individual feature to our models' behavior, we report two complementary measures: permutation feature importance (PFI) (Breiman, 2001) and Shapley additive explanations (SHAP) (Lundberg & Lee, 2017). While PFI measures importance to out-of-sample predictive performance, SHAP values measure the marginal impact, and its direction, of a feature's value on the prediction made by a model.

For a given model trained on the learning wave, its PFI is calculated by first establishing the model's baseline RMSE on the unpermuted validation set. Then, for each feature being evaluated, the feature's values are randomly shuffled in the validation set while keeping other features unchanged; the model's RMSE on the shuffled validation data is then measured. The degradation in performance (relative to the baseline) indicates that feature's importance. This process is repeated 30 times to quantify the uncertainty over shuffles of a feature's value.

SHAP values are calculated using version 0.42.1 of the SHAP package in Python.

## S11.2. Feature importance for OLS, RF, XGB, and MLP

In the manuscript, we show these feature importance measures for the E-Net, our best-performing model. Here, we include the same analyses for the other four models. Across all five models, communication is consistently the most important feature to out-of-sample predictive power. Beyond communication, models vary in their ranking of feature importance to predictive power, yet the impacts of various features on predicted values (as indicated by SHAP values) are, in general, directionally consistent. To contextualize the variation in feature importance, it is important to consider the model's baseline predictive performance, as the interpretation of a poorly performing model may not be of practical value. Moreover, different functional forms and algorithms may learn different mappings from features to outcomes, and some approaches may be more susceptible to overfitting than others.

In each feature importance figure below (Figures S11–S14), panel (A) shows each feature's PFI to out-of-sample prediction and error bars indicate 95% confidence intervals arising from the random parameter shuffles. Panel (B) displays SHAP values showing how each feature contributes to individual predictions, with color indicating feature value and position showing the magnitude and direction of the feature's contribution to predicted efficiency under punishment. Note that all models also used the control (no punishment) efficiency to inform predictions, although it is omitted from the figures for clarity as it was not a design space parameter.

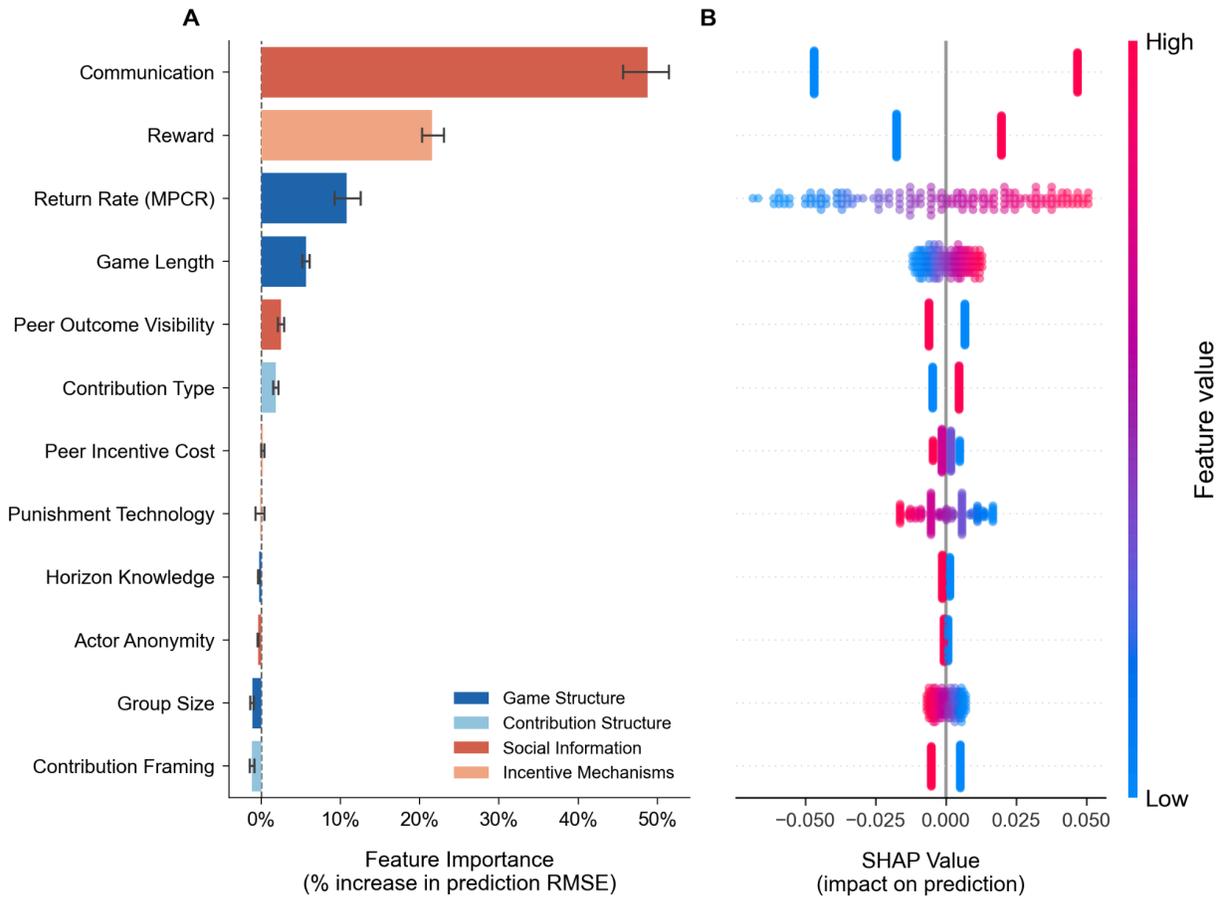

**Figure S11. Feature importance and model interpretation (OLS)**

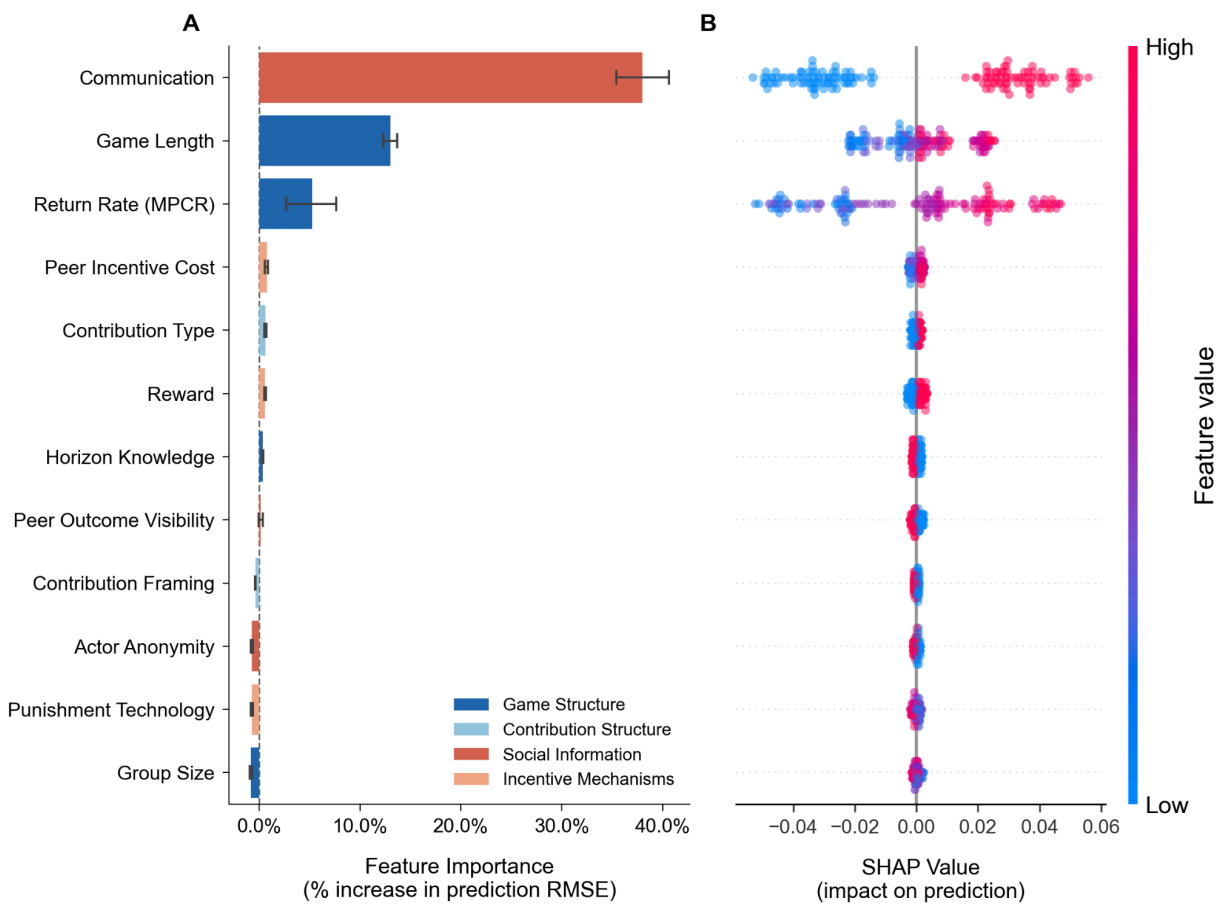

**Figure S12. Feature importance and model interpretation (Random Forest)**

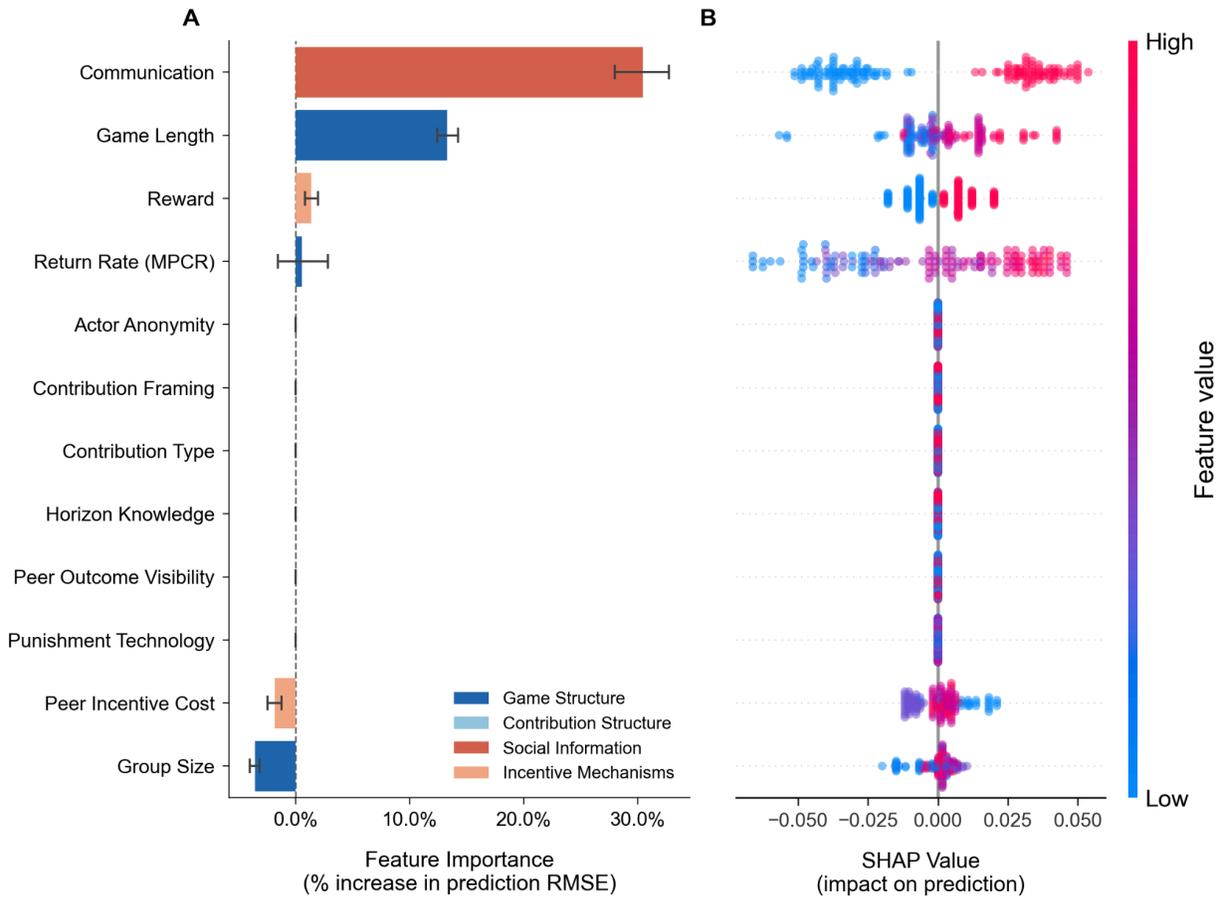

**Figure S13. Feature importance and model interpretation (XGBoost)**

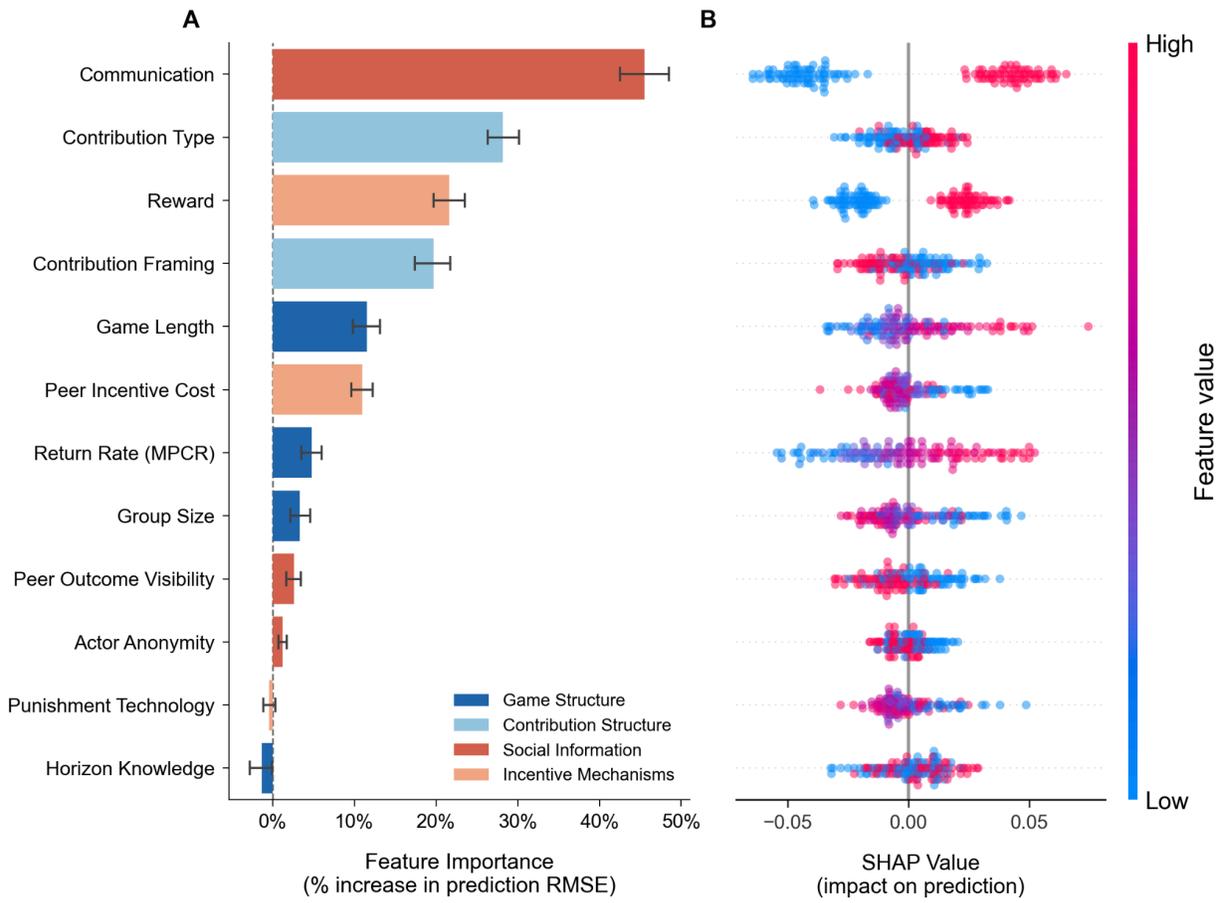

**Figure S14. Feature importance and model interpretation (Multilayer Perceptron)**